\documentclass[aps,prb,twocolumn,superscriptaddress]{revtex4-1}
\usepackage{graphicx}
\usepackage{epstopdf}
\usepackage{amsmath}
\usepackage{amsfonts}
\usepackage[usenames,dvipsnames]{color}

\begin{document}
\title{Unpaired Majorana modes on dislocations and string defects in Kitaev's honeycomb model}
\author{Olga Petrova}
\affiliation{Max Planck Institute for the Physics of Complex Systems, D-01187 Dresden, Germany
}
\author{Paula Mellado}
\affiliation{School of Engineering and Applied Sciences, 
	Adolfo Ib{\'a}{\~n}ez University,
	Santiago, Chile
}
\author{Oleg Tchernyshyov}
\affiliation{Department of Physics and Astronomy, 
	The Johns Hopkins University,
	Baltimore, Maryland 21218, USA
}

\begin{abstract}

We study the gapped phase of Kitaev's honeycomb model (a $Z_2$ spin liquid) on a lattice with topological defects. We find that some dislocations and string defects carry unpaired Majorana fermions. Physical excitations associated with these defects are (complex) fermion modes made out of two (real) Majorana fermions connected by a $Z_2$ gauge string. The quantum state of these modes is robust against local noise and can be changed by winding a $Z_2$ vortex around one of the dislocations. The exact solution respects gauge invariance and reveals a crucial role of the gauge field in the physics of Majorana modes. To facilitate these theoretical developments, we recast the degenerate perturbation theory for spins in the language of Majorana fermions. 

\end{abstract}

\maketitle

\section{Introduction}

In three dimensions all particles can be divided into two categories by their quantum statistics: bosons and fermions. The two  possibilities correspond to two one-dimensional representations of the permutation group of $N$ particles $S_N$. In two dimensions, the situation is richer because two particles can be exchanged clockwise or counterclockwise and the two exchange paths are topologically distinct. For this reason, particle exchange in two dimensions is called \emph{braiding} and exchange statistics is related to the braid group $B_N$, which has infinitely many one-dimensional representations. As a result, particle statistics can interpolate continuously between Bose and Fermi's, hence the name \textit{anyons}.\cite{Wilczek} When two Abelian anyons are exchanged, the system's wavefunction picks up a phase that is not restricted to integer multiples of $\pi$ as it is in three dimensions. 

Non-Abelian statistics corresponds to higher-dimensional representations of the braid group. It arises when the ground state of a system is degenerate and winding one particle around another amounts to a unitary transformation in the space of degenerate ground states. An anyon system is characterized by a set of \emph{fusion rules} that state the possible outcomes of fusing pairs of anyons. 

Since the result of braiding depends only on the topology of the braid, qubits made up of non-Abelian anyons are very stable with respect to any local perturbation. The anticipated applications in the field of quantum computing\cite{Kitaev2003} have been fueling the search for non-Abelian excitations, however, potentially physically realizable systems that give rise to them remain scarce. Here we discuss how adding topological lattice defects to the Abelian phase of the Kitaev honeycomb model \cite{Kitaev2006} can give rise to non-Abelian statistics. Such defects can be classified as twists, related to the $Z_2$ symmetry present in the model. It is worth noting that their presence in the system does not spoil its exact solvability, which allows us to explicitly demonstrate the crucial role of the gauge field in the physics of Majorana modes.

Twist defects are found in topologically ordered systems with a particular kind of symmetry: their fusion and braiding rules are invariant under the exchange of two distinct kinds of excitations. A twist is a point defect in two spatial dimensions (and a line defect in three) that alters the anyon type when an anyon is transported around it. An early precursor of the twist was the Alice string introduced by Schwarz,\cite{Schwarz} which induces electric charge conjugation in some gauge models. The possibility of anyon type exchange in a topological state was first suggested by Kitaev\cite{Kitaev2006} for the honeycomb model and studied by Barkeshli et al. \cite{BarkeshliWen2010,BarkeshliQi} in the context of fractional quantum Hall states. The first explicit construction of twist defects in a microscopic model was carried out by Bombin. \cite{Bombin} The name of the defect reflects the twisting of the underlying topological state up to a symmetry of the anyon model. \textcite{Barkeshli} suggested the term \emph{genon} to stress the connection between these defects and an increase in topological degeneracy.
\color{black}

In $Z_N$ rotor models (where $N=2$ case corresponds to the toric code \cite{Kitaev2003,Wen}) defined on a square lattice, excitations live at the ends of string operators, connecting diagonal plaquettes. Therefore, one defines two kinds of excitations, $e$ and $m$ topological charges, that can exist on odd and even plaquettes of a checkerboard lattice. Braiding $e$ and $m$ charges around one another gives rise to a phase factor, meaning that the two kinds of excitations are mutual Abelian anyons. Since the choice of even and odd plaquette type is arbitrary, the model is obviously symmetric under exchange of $e$ and $m$ anyons. A charge that winds around a $Z_2$ twist defect can be thought of as exchanging its type and the defect itself can be shown to behave as a non-Abelian anyon with quantum dimension $\sqrt{N}$. \cite{Bombin,You1}

A recent addition to the family of topologically ordered systems are topological nematic states. It is known that fractional quantum Hall states (FQH) can be realized in interacting lattice models with a non-trivial Chern number $C$. \cite{PhysRevLett.106.236802, PhysRevLett.106.236803, PhysRevLett.106.236804, PhysRevX.1.021014, FQH2011} For an integer $C>1$, such systems are equivalent to $C$ parallel FQH layers, so that translations of the lattice can be thought of as permutation of the layers. Lattice dislocations in such systems also constitute twist defects, where the symmetry in question exchanges layers.\cite{BarkeshliQi}

Realizations of twist defects are not limited to dislocations in lattice models. Other examples include edge states in domain walls between FQH regions gapped by two different means: for instance, by proximity to a superconductor and a ferromagnet, \cite{PhysRevX.2.041002, ClarkeAlicea,PhysRevB.87.035132} etc.

In this paper we present an explicit construction of non-Abelian quasiparticles---Majorana modes---in Kitaev's spin model on a honeycomb lattice.\cite{Kitaev2006} This spin model can be exactly solved by representing spins in terms of Majorana fermions living in the background of a static $Z_2$ gauge field. In the gapped phases of this model, low-energy excitations are $Z_2$ vortices, which come in two flavors living on alternating rows of hexagons of the lattice. A vortex of one flavor cannot be converted into another without creating or destroying additional quasiparticles. A lattice dislocation may act as a twist defect if moving a vortex around it returns the vortex to the wrong row of hexagons, thereby altering its flavor. The additional quasiparticle created in the process of conversion is a nonlocal fermion formed by two Majorana modes associated with the dislocation in question and with another dislocation elsewhere in the system. One pair of twists increases the degeneracy of the ground state by a factor 2. Unitary transformations in the Hilbert space of degenerate ground states can be achieved by braiding vortices around twists. We study the robustness of the Majorana modes to local perturbations such as an external magnetic field and demonstrate that the energy splittings induced in this way decay exponentially with the distance between dislocations. Furthermore, the interactions between Majorana modes are strongly directional: Majorana modes at two dislocations may not interact at all for certain relative positions. This directional effect has been previously noted by \textcite{Willans} who studied non-topological defects of the lattice such as vacancies in the same model. Some of our results have been reported in an earlier short communication.\cite{PhysRevB.88.140405}

The paper is organized as follows: first, we give an overview of the Kitaev honeycomb model in Section \ref{sec:model}. Since the focus of our work is on the gapped phase of the model, we make extensive use of high order perturbation theory. In order to simplify such calculations, we introduce a diagrammatic approach to perturbation theory in Section \ref{sec:diagrammatic-perturbation}, which may be of use for a wider range of problems. The technical details behind this method are given in Appendix A.

In this work, we consider two kinds of lattice dislocations: 8--2 (consisting of an octagon and a site with reduced coordination number 2) and 5--7 (composed of a pentagon and a heptagon). Both types may be \emph{trivial} (not result in unpaired Majorana modes) and \emph{twist} depending on their topological details. We start our discussion with 8--2 dislocations, which preserve the topology of link types in the lattice, in Section \ref{sec:82}. 5--7 dislocations, discussed in Section \ref{sec:57}, are made up of two disclinations which alter the orientation of the link flavors. We summarize our results in Section \ref{sec:discussion}.

\section{Kitaev's honeycomb model}
\label{sec:model}

\subsection{The spin model}

\begin{figure}
\centering
\includegraphics[width=0.9\columnwidth]{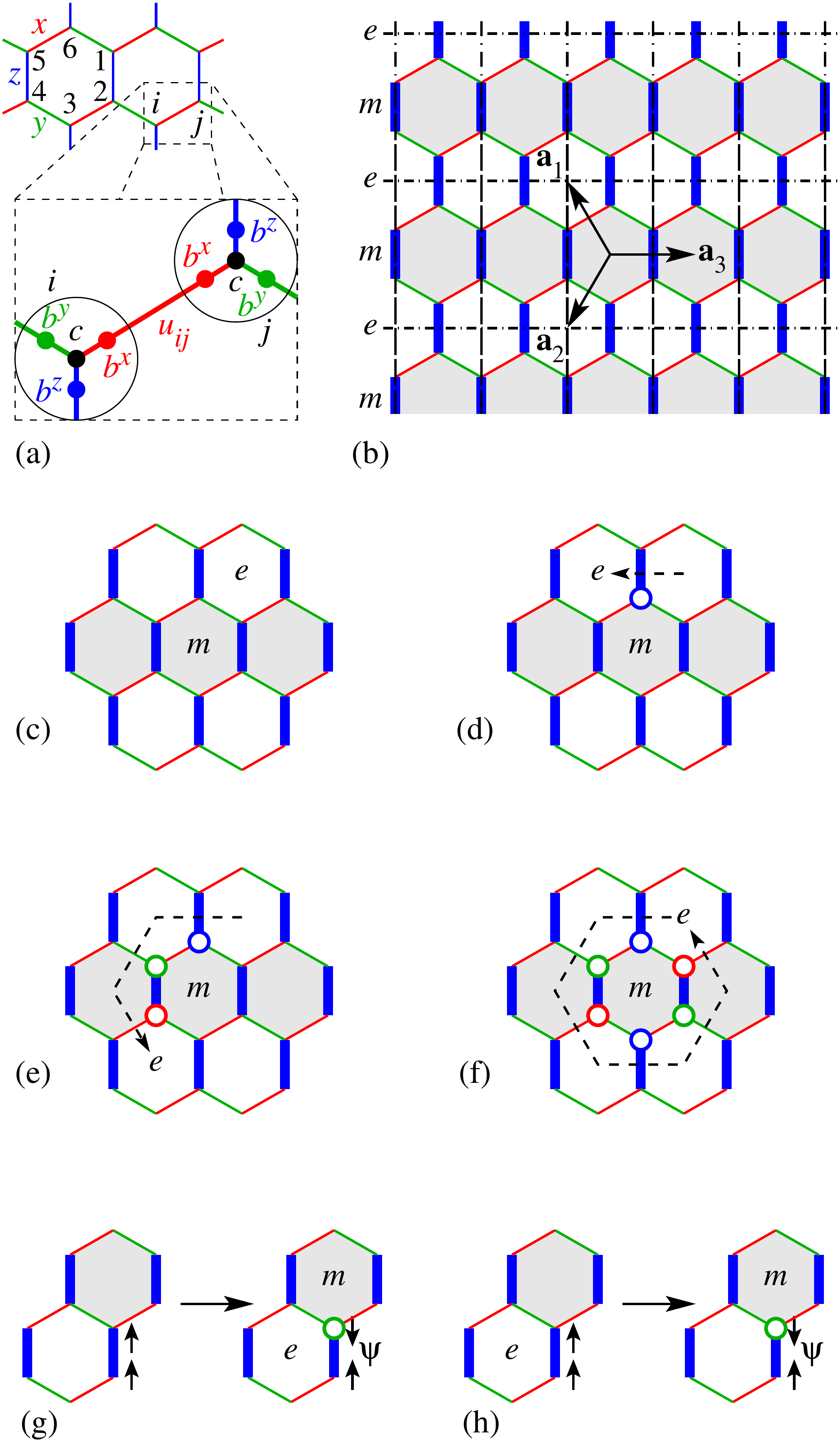}
\caption{(color) (a) Three types of bonds in the honeycomb lattice,  the conserved plaquette operator $W = \sigma^{x}_{1}\sigma^{y}_{2}\sigma^{z}_{3}\sigma^{x}_{4}\sigma^{y}_{5}\sigma^{z}_{6}$; lower panel: graphic representation of spins in terms of the Majorana operators $b^\alpha$ and $c$. (b) $e$ charges and $m$ fluxes in the toric code limit correspond to plaquettes where $W=-1$ on alternating blank and filled rows of the honeycomb lattice. Pairs of spins connected by the strong $z$ links become effective spins living on the dash-dotted bonds of the toric-code square lattice. $\mathbf a_i$ are lattice vectors. (c)--(f) Winding an $e$-type flux around an $m$-type flux multiplies the wavefunction by $-1$, indicating that $e$ and $m$ fluxes are mutual semions. (g)--(h) Local creation of an $e \times m$ vortex pair (g) and conversion of vortex flavor (h).  Open red, green, and blue circles denote the application of $\sigma^x$, $\sigma^y$, and $\sigma^z$ operators. }
\label{fig:intro}
\end{figure}

Kitaev's model has spins of length $S=1/2$ on sites of a honeycomb lattice, Fig.~\ref{fig:intro}(a). Adjacent spins are coupled through anisotropic two-spin interactions whose nature depends on the direction of the bond:
\begin{equation}
H = -J_x \sum_{x~\mathrm{links}} \sigma^{x}_{m}\sigma^{x}_{n}
	-J_y \sum_{y~\mathrm{links}} \sigma^{y}_{m}\sigma^{y}_{n}
	-J_z \sum_{z~\mathrm{links}} \sigma^{z}_{m}\sigma^{z}_{n},
\label{eq:HKitaev}
\end{equation}
where links are labeled as shown in Fig.~\ref{fig:intro}(a). Its solvability is related to the existence of integrals of motion, one for every hexagonal plaquette, 
\begin{equation}
W = \sigma_1^x \sigma_2^y \sigma_3^z \sigma_4^x \sigma_5^y \sigma_6^z, 
\label{eq:W-hexagon}
\end{equation}
in Fig.~\ref{fig:intro}(a). In the ground state, $W = +1$ everywhere.\cite{Kitaev2006} 

The Kitaev model has been extended to other lattices with coordination number 3 in two\cite{PhysRevLett.99.247203} and three\cite{PhysRevB.79.024426,arXiv:1403.3296,arXiv:1309.1171,PhysRevB.89.235102} dimensions.

\subsection{Majorana fermion representation}
\label{sec:majorana-fermion-representation}

Kitaev's exact solution is based on a representation of spin operators $\sigma_n^x$, $\sigma_n^y$, and $\sigma_n^z$ on a given site $n$ in terms of four Majorana fermions $b_n^x$, $b_n^y$, $b_n^z$, and $c_n$ that anticommute with one another and are normalized so that 
\begin{equation}
(b_n^x)^2 = (b_n^y)^2 = (b_n^z)^2 = c_n^2 = 1.
\label{eq:b-c-norm}
\end{equation}
The spin components are $\sigma_n^\alpha = i b_n^\alpha c_n$, Fig.~\ref{fig:intro}(a). The transformation from spin to fermion variables enlarges the dimension of the Hilbert space on each site from two to four. Physical states are eigenstates of the operator  
\begin{equation}
D_n \equiv b_n^x b_n^y b_n^z c_n 
\label{eq:D-def}
\end{equation}
with the eigenvalue $+1$, which guarantees the correct spin commutation relations, $[\sigma_m^\alpha, \sigma_n^\beta] = 2i \delta_{mn} \epsilon^{\alpha\beta\gamma} \sigma_n^\gamma$. 

The Hamiltonian expressed in terms of Majorana fermions reads 
\begin{equation}
H = i\sum_{\langle mn \rangle}J_{\alpha_{mn}} u_{mn}c_m c_n,
\label{eq:Hmajorana1}
\end{equation}
where $\langle mn \rangle$ denotes a nearest-neighbor bond connecting sites $m$ and $n$; spin component $\alpha = \alpha_{mn}$ depends on the orientation of the bond. The Hamiltonian contains explicitly the $c$ Majorana modes, whereas the $b$ modes are hidden in link variables $u_{mn} = -u_{nm} \equiv i b_m^\alpha b_n^\alpha$. The link variables are constants of motion that can be set to $u_{mn} = \pm 1$, thus reducing the Hamiltonian (\ref{eq:Hmajorana1}) to a quadratic form in $c$ Majorana fermions. It can then be diagonalized by an orthogonal transformation to a new set of Majorana modes, $c_m = \sum_n O_{mn} \gamma_n$, whose Hamiltonian is block-diagonal,
\begin{equation}
H =  \sum_n \frac{i \epsilon_n}{2}  \gamma_{2n-1} \gamma_{2n} 
	=  \sum_n \frac{\epsilon_n}{2} (\psi^\dagger_n \psi_n - \psi_n \psi^\dagger_n).
\label{eq:H-gamma-psi}
\end{equation}
Here $\epsilon_n \geq 0$ represents the excitation energy of a pair of Majorana modes $\{\gamma_{2n-1}, \, \gamma_{2n}\}$ that can be combined to form a complex fermion mode $\psi_n$,
\begin{equation}
\psi_n = \frac{\gamma_{2n-1} + i\gamma_{2n}}{2}, 
\quad 
\psi_n^\dagger = \frac{\gamma_{2n-1} - i\gamma_{2n}}{2}.
\end{equation}
The ground state of the Hamiltonian (\ref{eq:H-gamma-psi}) is the vacuum of the $\psi$ fermions annihilated by every operator $\psi_n$. It has the energy 
\begin{equation}
E = - \sum_n \epsilon_n/2. 
\label{eq:E-zero-point}
\end{equation}

The excitation spectrum of the Majorana modes $\{\epsilon_n\}$ is gapless in the thermodynamic limit if the coupling constants satisfy the triangle inequalities, $|J_x| + |J_y| > |J_z|$ and its permutations. Here we are interested in the gapped phases, where one of the coupling constant dominates, e.g., $|J_z| > |J_x| + |J_y|$. 

\subsection{$Z_2$ gauge symmetry}

Although link variables $u_{mn}$ are conserved quantities, they do not commute with the operators (\ref{eq:D-def}) constraining the physical states. In other words, they do not represent physically observable quantities. Assigning them a definite value $\pm 1$ is akin to fixing a gauge. Indeed, it is useful to view the constraint operators (\ref{eq:D-def}) as generators of a $Z_2$ gauge symmetry acting on the fermionic variables as follows: 
\begin{eqnarray}
b_n^\alpha &\mapsto& D_n^\dagger b_n^\alpha D_n = - b_n^\alpha, 
\nonumber\\
c_n &\mapsto& D_n^\dagger c_n D_n = - c_n, 
\label{eq:D-gauge}
\\
u_{mn} &\mapsto& D_n^\dagger u_{mn} D_n = - u_{mn}.
\nonumber 
\end{eqnarray}
Physical variables such as spins $\sigma_n^\alpha = i b_n^\alpha c_n$ are gauge invariant. In addition, a gauge-invariant quantity can be obtained by taking a product of link variables around a closed loop, $u_{12} u_{23} \ldots u_{n1}$, which could be interpreted as a $Z_2$ magnetic flux with values $\pm 1$. For a hexagonal plaquette, this product yields, up to a sign, the integral of motion $W$ defined in Eq.~(\ref{eq:W-hexagon}). 

\textcite{Kitaev2006} defined the $Z_2$ flux through a hexagonal plaquette [Fig.~\ref{fig:intro}(a)] in two ways:
\begin{subequations}
\begin{eqnarray}
W & \stackrel{?}{=} & \sigma_1^x \sigma_2^y \sigma_3^z \sigma_4^x \sigma_5^y \sigma_6^z, 
\label{eq:W-def-1}
\\
W & \stackrel{?}{=} & (\sigma_1^z \sigma_2^z)(\sigma_2^x \sigma_3^x) \ldots (\sigma_5^x \sigma_6^x)(\sigma_6^y \sigma_1^y).
\label{eq:W-def-2}
\end{eqnarray}
\end{subequations}
Although the two definitions are equivalent on a regular honeycomb lattice, this is not always the case in the presence of lattice disorder or on other three-coordinated lattices that can support the Kitaev model Hamiltonian. For instance, on a plaquette with an odd number of sites the two definitions differ by a factor of $\pm i$. It is therefore desirable to select a definition applicable to plaquettes of arbitrary shape. 

It seems reasonable to expect that $Z_2$ flux satisfies the following rules:  
\begin{enumerate}
\item For any plaquette, $W$ takes on one of the $Z_2$ values, $+1$ or $-1$. 
\label{item:rule-1}
\item For adjacent plaquettes 1 and 2, the flux through the combined plaquette 1+2 is the product of their individual fluxes, $W_{1+2} = W_1 W_2$.
\label{item:rule-2}
\end{enumerate}
Surprisingly, it does not seem possible in general to satisfy both rules simultaneously. Definition (\ref{eq:W-def-1}) satisfies rule (i). Rule (ii) is satisfied if link flavors are consistently oriented on every site, following the pattern $x$, $y$, $z$ as we go around a site counterclockwise. If some sites follow the opposite pattern $z$, $y$, $x$, rule (ii) is violated. Definition~(\ref{eq:W-def-2}) violates rule (i), giving $W=\pm i$ for plaquettes with an odd number of sites. However, it satisfies rule (ii). 

We view multiplicativity as the more basic property of $Z_2$ flux and therefore stick with Eq.~(\ref{eq:W-def-2}). We now must keep in mind that the flux on a plaquette with an odd perimeter depends on direction: if going clockwise yields $W = +i$ then going counterclockwise would yield $W = -i$. We shall see below that this quirky behavior makes sense. The energy of the system depends on fluxes through plaquettes with an even perimeter (where $W$ is real) but not on fluxes on plaquettes with an odd perimeter (where $W$ is imaginary). 

We thus define the $Z_2$ flux on a plaquette with sites $1, 2, \ldots, n$ on the boundary, going counterclockwise, as
\begin{equation}
W = (\sigma_1^{\alpha_{12}} \sigma_2^{\alpha_{12}}) (\sigma_2^{\alpha_{23}} \sigma_3^{\alpha_{23}}) \ldots 
	(\sigma_n^{\alpha_{n1}} \sigma_1^{\alpha_{n1}}).
\label{eq:W-def-sigma}
\end{equation}
After converting each link product to Majorana variables,
\[
\sigma_1^{\alpha_{12}} \sigma_2^{\alpha_{12}} 
	= (i b_1^{\alpha_{12}} c_1) (i b_2^{\alpha_{12}} c_2) 
	= - i u_{12} \, c_1 c_2,
\]
and after using the normalization condition $c_n^2 = 1$, we obtain the flux in terms of $Z_2$ gauge variables:
\begin{equation}
W = (-i)^n \, u_{12} u_{23} \ldots u_{n1},
\label{eq:W-def-u}
\end{equation}
which agrees with Eq.~(16) of Kitaev.\cite{Kitaev2006} 

\subsection{$Z_2$ magnetic vortices in the gapped phases}
\label{sec:majorana-fermion-representation-vortices}

We shall focus our attention on one of the gapped phases, where one of the coupling constants in Eq.~(\ref{eq:HKitaev}) dominates, e.g., $J_{z} > J_{x} + J_{y}$. We assume ferromagnetic couplings, $J_\alpha>0$, without loss of generality. The physics simplifies in the limit $J_{z} \gg J_x, \, J_y$, where low-energy states have parallel spins on strong ($z$) bonds. The ground state $|0\rangle$ is in the sector with $W=+1$ on all hexagonal plaquettes and with no fermions present, $\psi_n |0\rangle = 0$. 

Excitations come in two forms, fermions and $Z_2$ vortices. Fermion excitations $\psi$ are associated with breaking the alignment of spins on strong bonds and thus have a high energy cost of approximately $2J_z$, so we shall refer to them as high-energy fermions. Low-energy excitations are $Z_2$ vortices, $W = -1$, with  energy $J_x^2 J_y^2/8J_z^3$. \cite{Kitaev2006} The effective Hamiltonian in this subspace turns out to be the toric code,\cite{Kitaev2003,Wen} with effective spins $\tau_{mn}^z = \sigma_m^z = \sigma_n^z$ living on links of a rectangular lattice, Fig.~\ref{fig:intro}(b).  Crucially, vortex excitations come in two flavors---$e$ and $m$---depending on the plaquette. A honeycomb plaquette centered on a vertex (plaquette) of the rectangular toric-code lattice may host an $e$ ($m$) vortex. Thus $e$ and $m$ vortices live in alternating rows, Fig.~\ref{fig:intro}(b). 

In the toric code, $e$ and $m$ particles are mutual semions: the wavefunction acquires a minus sign when a particle of one type winds around a particle of the other type. The same is true of the $e$ and $m$ vortices in the honeycomb spin model, Fig.~\ref{fig:intro}(c)--(f). Here the winding is accomplished by the application of six $\sigma$ operators, whose product equals the flux (\ref{eq:W-hexagon}) on the central plaquette, $W=-1$. Exchanging two $e\times m$ pairs also produces a minus sign, pointing to the fermionic nature of the composite $e\times m$ particle. In the unperturbed toric code model there is no way to have an odd number of $e\times m$ pairs. The underlying reason for this is that the parity of the total number of fermions in the system should be conserved. Much of the toric code description carries over to the vortices of the honeycomb model with different flavors. However, unlike in the toric code, there is nothing that forbids us from creating a vortex pair in adjacent rows of the lattice pictured in Fig.~\ref{fig:intro}(g), seemingly breaking the fermion parity conservation. The mystery is solved by examining the process in terms of the honeycomb spins: the $e \times m$ vortex pair in adjacent rows is created via the application of an operator $\sigma_n^x$ or $\sigma_n^y$, which misalignes a pair of spins connected by a strong $z$ bond. It follows that creating a $e\times m$ vortex pair is accompanied by creation or annihilation of a fermion $\psi$ with a high energy cost of $2J_z$. Such processes, as well as the local conversion of the vortex flavor [Fig.~\ref{fig:intro}(h)], are effectively forbidden at low energies.

\begin{figure}
\includegraphics[width=0.99\columnwidth]{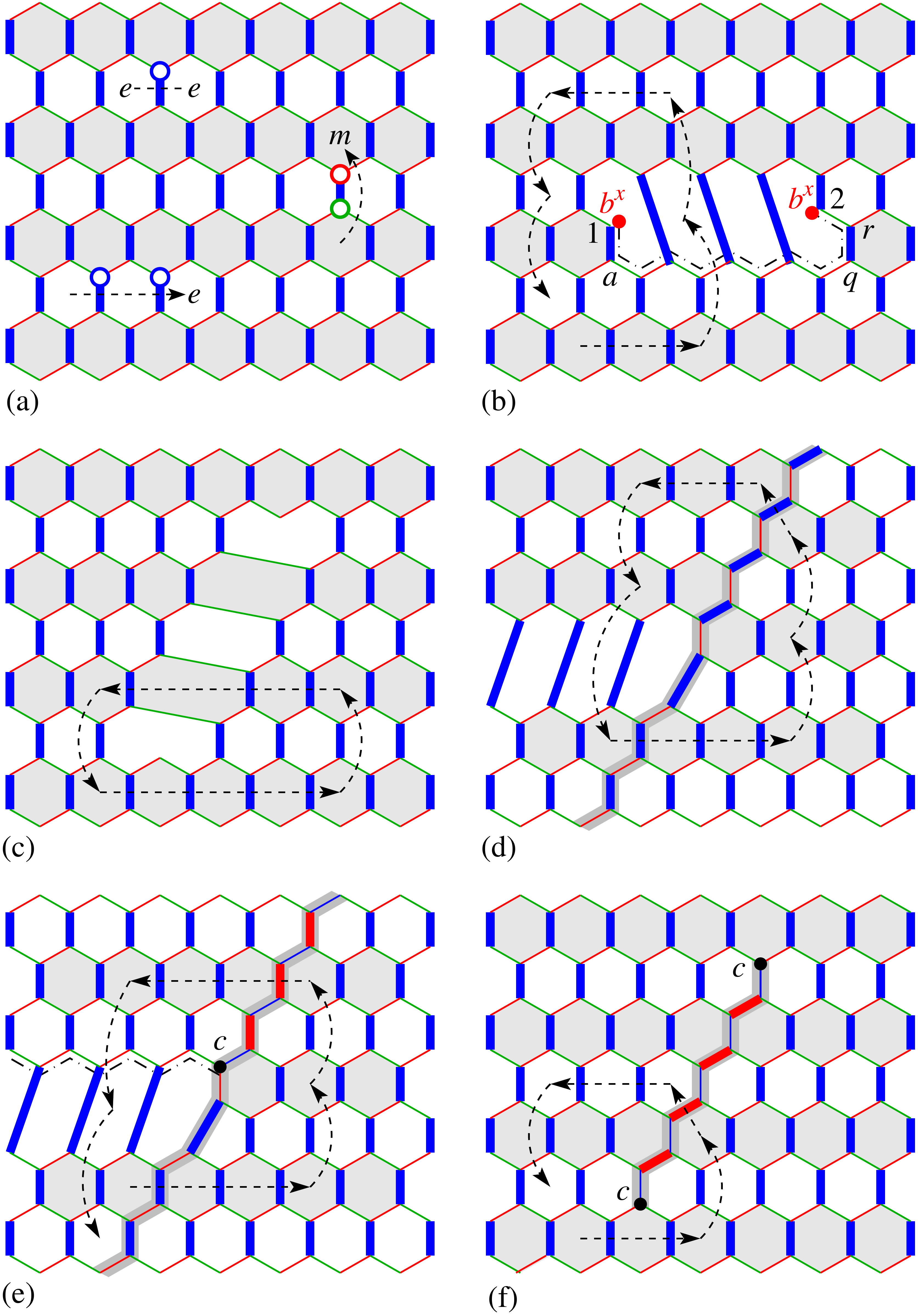}
\caption{(color) (a) Low-energy excitations in the gapped phase of the model. Vortices are located at the ends of dashed lines, representing string operators that flip the value of $W_p$ when they cross into, or out of, a plaquette. Open red, green, and blue circles indicate $\sigma^x$, $\sigma^y$, and $\sigma^z$ operators. (b)--(c) A pair of 8--2 dislocations: twists with $\mathbf B = \pm \mathbf a_1$ (b) and trivial with $\mathbf B = \pm \mathbf a_3$ (c). (d)--(e) A pair of 5--7 dislocations with $\mathbf B = \pm \mathbf a_2$: trivial (d) and twists (e). (f) A line of links with altered dimerization with twists at the ends. Dashed lines are flux path. Dash-dotted lines are branch cuts. Filled circles are unpaired Majorana modes.}
\label{fig:fig2}
\end{figure}

Consequently the low-energy processes are restricted to (a) creating and annihilating two vortices in the same row of the honeycomb lattice; (b) shifting a vortex in its row; (c) a vortex hopping to the next-nearest row, Fig.~\ref{fig:fig2}(a). The first two are accomplished by acting with an operator $\sigma_n^z$; the third by applying $\sigma_m^{x} \sigma_n^{y}$ or $\sigma_m^{y} \sigma_n^{x}$ on a strong bond $\langle mn \rangle$. 

\subsection{Fermion parity}
\label{sec:parity}

\begin{figure}
\includegraphics[width=0.99\columnwidth]{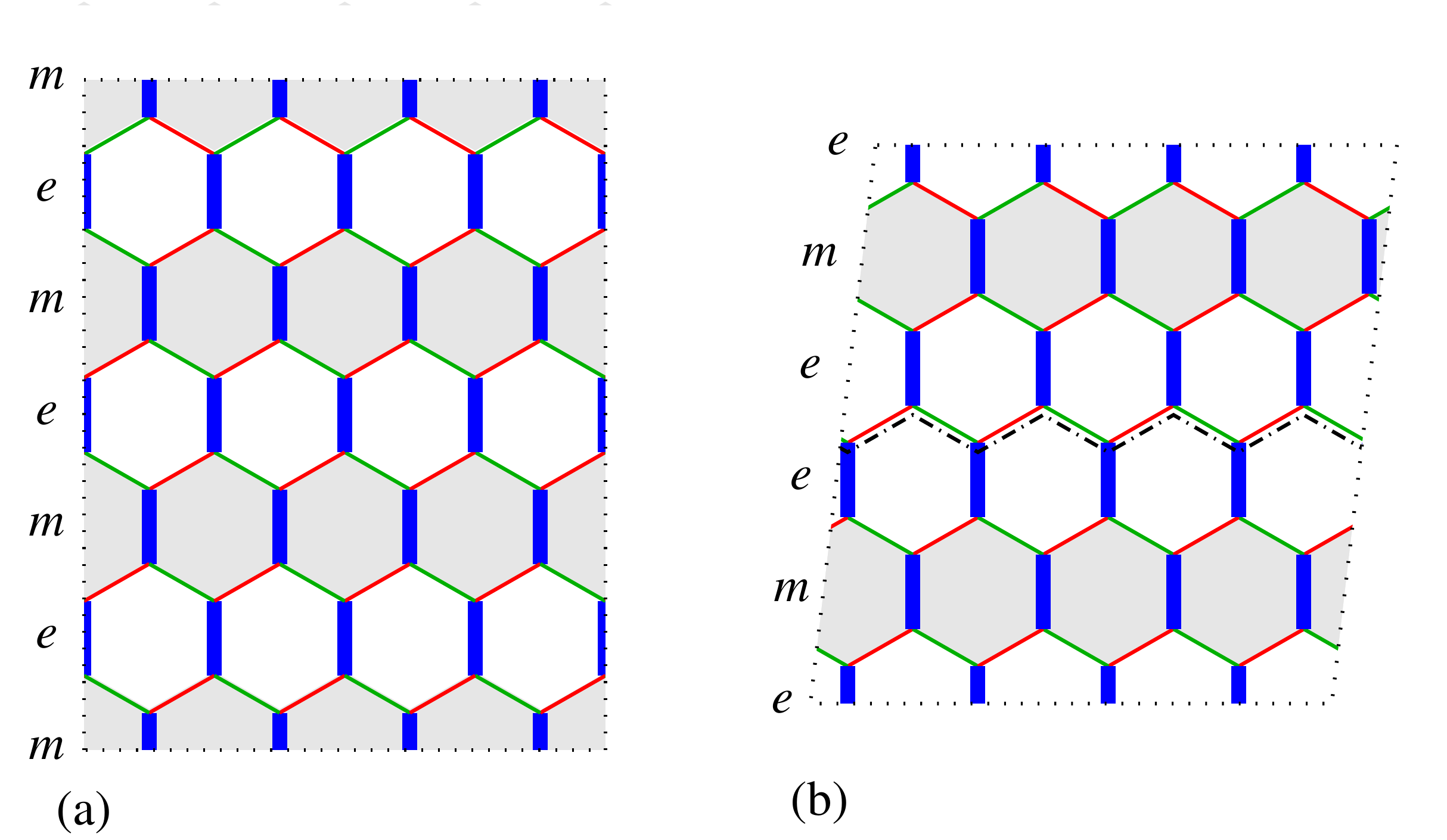}
\caption{The system on tori: opposite edges of the parallelograms are identified. (a) A torus with 6 rows of hexagons. (b) A torus with 5 rows of hexagons. The dotted-dashed line is a cut separating two rows of $e$ hexagons.}
\label{eq:fig-torus}
\end{figure}

Physical variables such as spin $\sigma^\alpha_n$ are bilinear in the Majorana operators, $\sigma^\alpha_n = i b^\alpha_n c_n$. This principle also works at the level of elementary excitations: a local operation always creates and destroys fermions in pairs. The conservation of fermionic parity has been tied to $Z_2$ gauge invariance by \textcite{Pedrocchi}. They have shown that, in a given flux sector, parity of the physical fermions $\{\gamma\}$ or $\{\psi\}$ diagonalizing the energy (\ref{eq:H-gamma-psi}) remains fixed. Their proof was general and independent of the system's Hamiltonian. Here we adopt it to the specific case of the gapped phase. Narrowing the focus allows us to compare fermionic parity in different flux sectors. This is of interest to us because a pair of fluxes of $e \times m$ types is also a fermion. It is reasonable to expect that gauge invariance translates into the conservation of the \emph{net} fermion parity, which counts both $\psi$ and $e \times m$ fermions. We demonstrate this explicitly for the case of simple topology: a torus with an even number of rows, Fig.~\ref{eq:fig-torus}(a). This will set the stage for a discussion of Majorana modes on twist dislocations, whose parity contributes to the net parity budget. The story has an interesting twist (so to speak) when the torus has an odd number of rows, Fig.~\ref{eq:fig-torus}(b). 

\subsubsection{Parity of $\psi$ fermions}

Deeply in the gapped phase where $z$ links dominate we may drop the $x$ and $y$ terms in the Hamiltonian as a starting point: 
\[
H_0 = -J_z \sum_{z~\mathrm{links}} \sigma^{z}_{m}\sigma^{z}_{n} = J_z \sum_{z~\mathrm{links}} i u_{mn} c_m c_n.
\]
Complex fermionic eigenmodes of this Hamiltonian live on strong links: 
\begin{equation}
\psi_{mn} = \frac{c_m + i u_{mn} c_n}{2},
\quad
\psi_{mn}^\dagger = \frac{c_m - i u_{mn} c_n}{2}.
\end{equation}
The Hamiltonian translates into 
\[
H_0 = J_z \sum_{z~\mathrm{links}} (\psi_{mn}^\dagger \psi_{mn} - \psi_{mn} \psi_{mn}^\dagger),
\]
showing that each complex fermion has positive excitation energy $+2 J_z$. The parity of the $\psi$ fermions is
\begin{eqnarray}
\pi_\psi &=& \prod_{z~\mathrm{links}} (\psi_{mn} \psi_{mn}^\dagger - \psi_{mn}^\dagger \psi_{mn})
\nonumber\\
	&=& \prod_{z~\mathrm{links}} (-i u_{mn} c_m c_n) 
	= \prod_{z~\mathrm{links}} \sigma_m^z \sigma_n^z.
\label{eq:pi-psi}
\end{eqnarray}

Restoring the weak $x$ and $y$ terms in the Hamiltonian will mix these modes. The parity of the new modes $\Psi'$ will be related to $\pi_\psi$ by the determinant of the linear transformation,\cite{Pedrocchi} which can only be $+1$ or $-1$. If the change of variables were small, the transformation would be close to identity and so its determinant could only be $+1$. However, the starting point is highly degenerate: all $\psi$ fermions have the same energy $2J_z$, so they are thoroughly mixed. 

Fortunately, the presence of a large energy gap means that $\psi$ operators (energy $J_z$) are mixed among themselves and $\psi^\dagger$ operators (energy $-J_z$) are mixed separately. Therefore, the transformation matrix is (approximately) block-diagonal: symbolically,  
\[
\left( 
	\begin{array}{l}
		\Psi\\
		\Psi^\dagger
	\end{array}
\right)
\mapsto 
\left( 
	\begin{array}{ll}
		\mathcal U & 0\\
		0 & \mathcal U^*
	\end{array}
\right)
\left( 
	\begin{array}{l}
		\Psi\\
		\Psi^\dagger
	\end{array}
\right),
\]
where $\mathcal U^*$ is the complex conjugate of the unitary transformation matrix $\mathcal U$. The determinant of this transformation is
\[
\det{\mathcal U} \det{\mathcal U^*} 
	= \det{\mathcal U} \det{\mathcal U^\dagger} 
	= \det{\mathcal U} \det{\mathcal U^{-1}} = +1.
\]
Thus the parity of the complex fermions $\pi_\psi$ is unchanged by the transformation, so we may safely use Eq~(\ref{eq:pi-psi}) to represent the parity of the complex $\psi$ fermions in the gapped phase. 

\subsubsection{Contribution of fluxes}
\label{sec:contribution-of-fluxes}

A physical state is an eigenstate of every gauge transformation $D_n$ (\ref{eq:D-def}) with the eigenvalue $+1$. An eigenstate $|\psi_u\rangle$ of the free Majorana Hamiltonian (\ref{eq:Hmajorana1}) with fixed $Z_2$ gauge field variables $u_{mn}$ is not gauge invariant and thus does not belong to the physical subspace. A physical state $|\psi_w\rangle$ can be obtained from it by summing over all possible gauge transformations of $|\psi_u\rangle$:\cite{Kitaev2006}
\begin{equation}
|\psi_w\rangle=\prod_{n=1}^{2N}\frac{1+D_n}{2}|\psi_u\rangle.
\label{eq:physpace}
\end{equation}
$|\psi_w\rangle$ is invariant under all $D_n$, and, consequently, under their product $\prod_n D_n$. The latter leaves the individual terms in the superposition (\ref{eq:physpace}) invariant. Therefore, one can symmetrize a state  $|\psi_u\rangle$ over all possible gauge transformations and obtain a physical state when $|\psi_u\rangle$ is an eigenstate of  $\prod_n D_n$ with the eigenvalue $+1$. States with eigenvalue $-1$ will be eliminated by $\prod_{n=1}^{2N}\frac{1+D_n}{2}$, so one may think of that operator as a projector onto the physical subspace.\cite{Pedrocchi}

We factorize each $D_n$ into the product of commuting operators $b^x_n b^y_n$ and $b^z_n c_n$ and rearrange them: 
\[
\prod_n D_n = \prod_m b^x_m b^y_m \prod_n b^z_n c_n. 
\]
In the second product on the right-hand side, we pair sites connected by strong bonds: 
\[
(b^z_m c_m) (b^z_n c_n) = - (b^z_m b^z_n) (c_m c_n) = i u_{mn} c_m c_n,
\]
whence
\[
\prod_n D_n = (-1)^{N/2} \pi_\psi \prod_m b^x_m b^y_m.
\]
where $N/2$ is the number of $z$ bonds in a system with $N$ sites.

Next we rearrange the $b^x$ and $b^y$ operators so as to form a product of alternating $x$ and $y$ links arranged in horizontal rows: 
\[
\prod_n D_n = \pi_\psi \prod_\mathrm{rows} (-1)(-iu_{12})(-i u_{23}) \ldots (-iu_{L1}), 
\]
where $L$ is the number of sites in a horizontal row. On a torus with an even number of rows, Fig.~\ref{eq:fig-torus}(a), the factors of $-1$ contributed by every row cancel out. The product of $u$ variables for two adjacent rows of links yields the net flux $W$ through the hexagons between them. Depending on how we pair the link rows, we end up with the net flux of hexagons of $e$ or $m$ type:
\[
\prod_n D_n = \pi_\psi W_e = \pi_\psi W_m. 
\]
In fact, they should be the same, $W_e = W_m$, because their product gives the net flux through the hexagons of the torus, $W_e W_m = +1$. 

\subsubsection{Net fermion parity}
\label{sec:parity-fermion}

Because an $e \times m$ fermion is a combination of an $e$ flux and an $m$ flux, $W_m = W_e$ coincide with the parity of the $e \times m$ fermions $\pi_{e \times m}$. We thus obtain the anticipated result: 
\begin{equation}
\pi_\psi \pi_{e\times m} = \prod_n D_n = +1.
\label{parity-conservation}
\end{equation}

The same conclusion can be arrived at much faster by using the spin representations for flux (\ref{eq:W-hexagon}) and fermion parity (\ref{eq:pi-psi}).  The product of all $m$ fluxes and $\psi$ fermion parity is
\[
W_m \pi_\psi = 1,
\]
thus, the net parity of physical fermions, counting both high-energy excitations $\psi$ and low-energy composite fermions $e\times m$, is 1. This is always the case for states described in terms of spins, whereas in the Majorana fermion representation $\pi_\psi \pi_{e\times m} =\pm1$. Only states that yield $+1$ correspond to the physical subspace.

\section{Transmutation of vortex flavor}

\subsection{Torus with a twist}
\label{sec:parity-fermion-twist}

The torus in Fig.~\ref{eq:fig-torus}(b) has an odd number of rows. As a result, it is impossible to globally partition hexagons into alternating $e$ and $m$ rows: there is a mismatch with two adjacent rows of the same flavor. The defect line, which we will call the cut, goes around the torus and can be deformed and moved around but cannot be eliminated. 

We can see that the cut must play a role in the overall budget of total fermion parity as follows. Starting in a ground state, we create a pair of fluxes of the same type, say $m$, and move one of them following a loop around the torus. As the flux crosses the cut, its type changes to $e$. As the $e$ flux returns to its $m$ partner, the two can be viewed as an $e \times m$ fermion. If the net fermion parity is conserved, the change of parity $\pi_{e \times m}$ must be compensated by another term. If the flux was moved gently, without exciting high-energy $\psi$ fermions, the compensating factor should be somehow related to the cut. 

To see how this issue is resolved, we compute the product $\pi_{e \times m} \pi_\psi$. Again, we do so by using spin variables, leading to the following result:
\[
W_m \pi_\psi = \prod_\mathrm{cut}\sigma_n^z,
\]
with the product of spins taken along the cut in Fig.~\ref{eq:fig-torus}(b). This product looks like a Wilson loop operator,\cite{Kitaev2006} and indeed it is. By rewriting $\sigma^z = - i \sigma^x \sigma^y$ and expressing the spin variables in terms of Majorana operators, we find that 
\[
-\prod_\mathrm{cut}\sigma_n^z = \prod_\mathrm{cut} (-iu_{mn}) \equiv W_\mathrm{cut}.
\]
The right-hand side is the global $Z_2$ flux piercing the vertical loop. 

Finally, after using the identity $W_m = W_e = \pi_{e \times m}$, we obtain the conservation law for a torus with an odd number of rows: 
\begin{equation}
\pi_\psi \pi_{e \times m} W_\mathrm{cut} = - 1.
\label{parity-conservation-twist}
\end{equation}
Moving an $m$ flux across the cut---at low energies (Sec.~\ref{sec:majorana-fermion-representation-vortices})---converts it into an $e$ flux, thereby altering the fermion parity. This is still consistent with Eq.~(\ref{parity-conservation-twist}) because a flux moving across the cut alters its Wilson-loop operator $W_\mathrm{cut}$. 

\subsection{Introducing lattice dislocations}

Vortices are created in pairs on neighboring hexagons and can then be brought further apart by flipping values of $W_p$ on plaquettes along the way. In Fig.~\ref{fig:fig2} such successive operations are depicted as dashed lines with vortices at the ends. A closed loop in Fig.~\ref{fig:fig2}(c) can then be thought of as creating a pair of $m$ vortices out of the vacuum and then one of them completing a loop via the low-energy movements, returning to the starting point. Since there are no lattice defects present, the vortex is able to return  to the same row and can then be annihilated with its partner to bring the system back to its original ground state. As can be seen from Fig.~\ref{fig:fig2}(b), (e), if the loop encloses a dislocation, the vortex may return to an adjacent $e$ row, in which case we end up with a composite $e\times m$ fermion. 

To gain a better understanding of the vortex type transmutation that took place, let us see what happens to the degrees of freedom when a dislocation defect is introduced into a system. First, consider the unperturbed model. In a system with $2N$ spins, there are $N$ plaquettes and $N$ strong bonds, giving rise to $N$ fluxes and $N$ fermion modes ($\psi$) respectively. However, not all of these degrees of freedom are independent. First, the net flux through is trivial, hence $W_m W_e = 1$; second, $\pi_\psi \pi_{e\times m}=+1$, Eq.~(\ref{parity-conservation}). This reduces the number of independent qubits to $N+N-2$, whereas the total number should be equal to the number of spins $2N$. The 2 remaining qubits correspond to closed string operators enclosing non-contractible loops winding around the torus, which give rise to the 4-fold topological degeneracy.

Let us introduce a dislocation pair into the system, so that the two dislocations are $\ell-1$ hexagons apart ($\ell=3$ in Fig.~\ref{fig:fig2}(b)). In the process, we remove $2\ell$ sites, $\ell$ strong bonds, and convert $2(\ell+1)$ hexagons into $\ell+1$ plaquettes ($\ell-1$ hexagons and 2 octagons). In other words, we are left with $N-\ell$ $\psi$ fermion modes and $N-\ell-1$ fluxes. Additionally, there are two global strings around the non-trivial loops of the torus. The number of independent qubits is reduced by 2 because we still have the relation $W_m W_e = 1$ and, similarly to the torus with an odd length, there is a constraint involving the fermion parities $\pi_\psi$ and $\pi_{e \times m}$ and a cut from one dislocation to another, Eq.~(\ref{eq:pi-psi-em-12}). If we now compare the number of degrees of freedom we have counted so far to the number of sites ($2N-2\ell$), we shall see that we are missing one qubit. This mode, a nonlocal fermion, is divided between the two dislocation cores. Along the same lines we find that adding $n$ dislocation pairs gives rise to 2 qubits associated with the non-contractible loops around the torus and $n$ qubits associated with the Majorana modes of the dislocations.

It is interesting to note that the ground-state degeneracy on a torus with $n \geq 1$ dislocation pairs is not $2^{n+2}$, as one might infer from the count of qubits, but is $2^{n+1}$, as pointed out by You and Wen.\cite{You1} In other words, adding the first pair of twist dislocations does not alter the 4-fold topological degeneracy, whereas every additional pair will increase the degeneracy by a factor of 2. This can be understood in the following manner. If we start with a flux-free ground state and want to change the parity of the nonlocal fermion mode, the way to do so would be to create a pair of fluxes (say, $e$ and $e$) and wind one of them around a dislocation. However, the resultant pair of $e\times m$ fluxes cannot be annihilated to bring the system back to its ground state as this would require the creation of a high-energy bond fermion. The presence of additional dislocation pairs allows us to wind one of the fluxes around another twist, thereby changing the flux's type once more, and annihilate it with its counterpart. Therefore, each additional dislocation pair will contribute a factor of 2 to the system's topological degeneracy, pointing to a quantum dimension of $\sqrt{2}$ associated with each twist defect.

As suggested by the quantum dimension of $\sqrt{2}$, we can fix the apparent non-conservation of fermion parity by associating a Majorana zero mode $\beta$ with the dislocation defect. Majorana modes of two dislocations can be combined to form a nonlocal complex fermion $\Psi = (\beta_1 + i \beta _2)/2$. A vortex winding around one of the dislocations alters the fermion number $\Psi^\dagger \Psi$ to compensate for the creation of the composite $e \times m$ fermion. In what follows we make an explicit construction of Majorana modes on dislocations in Kitaev's honeycomb model.

A typical dislocation in graphene \cite{carpio2008dislocations, NatMater.9.792, yazyev2010topological, cortijo2007effects} is a composite object that consists of two disclinations with angles $+\pi/3$ and $-\pi/3$, containing at their cores plaquettes with 5 and 7 sites. (Hence the name: a 5--7 dislocation.) Being a combination of two disclinations, it alters the topology of link labels, creating a string defect on which bonds of two flavors have altered orientations, the upper half of the shaded line in Fig.~\ref{fig:fig2}(d). The simplest dislocation that preserves the topology of bond labels consists of an octagonal plaquette (a $-2\pi/3$ disclination) and a site with reduced coordination number 2 (a $+2\pi/3$ disclination), or 8--2 for brevity, Fig.~\ref{fig:fig2}(b) and (c). It can be synthesized in Kitaev's model by quenching a line of sites with a strong magnetic field as discussed in Section \ref{sec:synthetic}. We first discuss the more straightforward case of 8--2 dislocations.

To act as a transformer of vortex flavor, a dislocation must have a Burgers vector $\mathbf B$ connecting plaquettes of different types, e.g., $\mathbf B = \pm \mathbf a_1$ or $\pm \mathbf a_2$ in the gapped phase with strong $z$ bonds. Fig.~\ref{fig:fig2}(b) shows a dislocation with $\mathbf B = \mathbf a_1$. It can be seen that a vortex winding around this dislocation via low-energy moves alters its flavor. Following Bombin\cite{Bombin} and others,\cite{You1, Barkeshli} we refer to such dislocations as \emph{twists}. Fig.~\ref{fig:fig2}(c) shows a dislocation with $\mathbf B = -\mathbf a_3$, which preserves the vortex type and is in this sense \emph{trivial}.

\section{8--2 dislocations}
\label{sec:82} 

\subsection{Twist dislocations} 
\label{sec:twist-dislocations}

As can be seen in Fig.~\ref{fig:fig2}(b), the presence of a $\mathbf B = \mathbf a_1$ dislocation makes it impossible to partition the lattice into plaquettes of $e$ and $m$ flavors globally. Any locally consistent partition has a branch cut connecting two dislocation cores. An $e$ vortex crossing the branch cut turns into an $m$ vortex and vice versa.  

Because site 1 at the cusp of the octagonal core in Fig.~\ref{fig:fig2}(b) is missing a weak $x$ bond, its Majorana fermion $b_1^x$ is unpaired. To form a zero-energy (complex) fermion mode $\Psi$, we can combine $b_1^x$ with a dangling Majorana mode of another twist dislocation, e.g., $b_2^x$ in Fig.~\ref{fig:fig2}(b). The naive recipe, 
\begin{equation}
\Psi \stackrel{?}{=} \frac{b_1^x + i b_2^x}{2}, 
\quad 
\Psi^\dagger \stackrel{?}{=} \frac{b_1^x - i b_2^x}{2}, 
\end{equation}
does not work: the fermion parity $\pi_{12} = \Psi \Psi^\dagger - \Psi^\dagger \Psi = i b_1^x b_2^x$ is not a physical quantity because it is not gauge-invariant (odd under both $D_1$ and $D_2$). This problem can be fixed by adding a gauge string factor,\cite{Kitaev2006} 
\begin{equation}
U_{12} = u_{1a} u_{ab} \ldots u_{qr} u_{r2}, 
\label{eq:U12}
\end{equation}
where $1ab \ldots qr2$ is a path connecting dislocation cores 1 and 2 as depicted  in Fig.~\ref{fig:fig2}(b). We have
\begin{equation}
\Psi = \frac{b_1^x + i U_{12} b_2^x}{2}, 
\quad 
\Psi^\dagger = \frac{b_1^x - i U_{12} b_2^x}{2}.
\end{equation}
The fermion parity 
\begin{equation}
\pi_{12} = \Psi^\dagger \Psi - \Psi \Psi^\dagger = i U_{12} b_1^x b_2^x 
\label{eq:pi12}
\end{equation}
is now gauge invariant and can be expressed as a product of spin operators along the string,

\begin{eqnarray}
\pi_{12} 
	&=& \sigma_1^y \sigma_a^x \ldots \sigma_q^y \sigma_r^x \sigma_2^z
\label{eq:pi12-spins}
\\
	&=& \sigma_1^x (i \sigma_1^{\alpha_{1a}} \sigma_a^{\alpha_{1a}}) 
		\ldots
		(i \sigma_q^{\alpha_{qr}} \sigma_r^{\alpha_{qr}})
		(i \sigma_r^{\alpha_{r2}} \sigma_2^{\alpha_{r2}}) i \sigma_2^x.
\nonumber
\end{eqnarray}

Like a branch cut, a string does not have a well-defined position; only its ends are fixed at dislocation cores. 

We can now see that the state of this fermion mode is altered when a flux winds around either of the dislocations. When the path of the flux crosses the string $1ab \ldots r2$, the link variable $u_{mn}$ at their crossing changes sign. This alters the sign of the gauge string (\ref{eq:U12}) and thereby changes parity (\ref{eq:pi12}). We have thus established that the variable $b_m^\alpha$ of an octagon cusp missing a weak bond $\alpha$ is the Majorana mode associated with a twist dislocation. Together with a Majorana fermion of another dislocation, it forms a zero-energy mode whose quantum state can be changed by winding a flux around one of the dislocations.

\begin{figure}
\includegraphics[width=0.9\columnwidth]{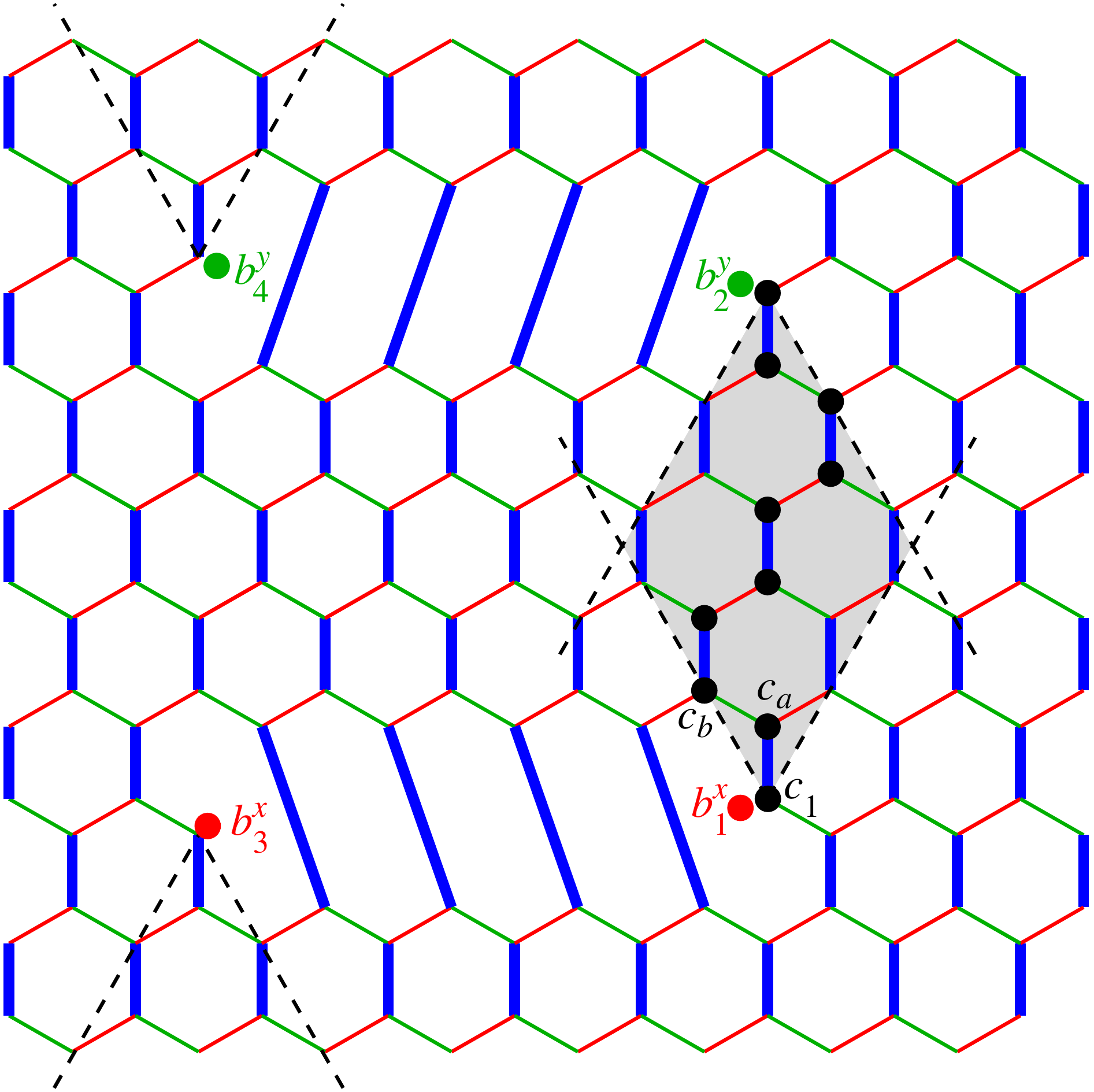}
\caption{Effective interaction between dangling Majorana modes $b_{1x}$ and $b_{2y}$ (colored dots) is generated by integrating out $c$ fermions along paths within the shaded area formed by the overlapping 60-degree wedges (dashed lines). Black dots indicate a sample path. Dangling modes $b_{3x}$ and $b_{4y}$ do not interact with each other because their wedges do not overlap.}
\label{fig:wedges}
\end{figure}

Having established the nature of the Majorana modes at twist dislocations, we can estimate their tolerance to local perturbations. In the presence of a magnetic field $\mathbf h = (h_x,h_y,h_z)$, the four dangling Majorana modes $b_m^\alpha$ in Fig.~\ref{fig:wedges} are coupled to the rest of the system by the Zeeman term $- h_\alpha \sigma_m^\alpha = - i h_\alpha b_m^\alpha c_m$. This coupling may lift the degeneracy of the zero mode and induce its time evolution, an undesirable effect, especially if the field is noise. We shall see that the splitting decays exponentially with the distance between dislocations. The adverse effects of local noise can be suppressed by keeping dislocations sufficiently far apart.

To compute the splitting of the zero mode, we integrate out the high-energy $c$ modes as explained in Sec.~\ref{sec:preliminaries-perturbation-example-1}. Consider first three sites in the vicinity of dislocation core 1, namely 1, $a$, and $b$. Their Majorana modes $b_{x1}$, $c_1$, $c_a$, and $c_b$ are coupled to one another as follows: 
\begin{eqnarray}
H &=& - h_x \sigma_1^x - J_z \sigma_1^z \sigma_a^z - J_y \sigma_a^y \sigma_b^y
\nonumber\\
&=& - h_x \, i b_1^x c_1 + J_z u_{1a} \, i c_1 c_a + J_y u_{ab} \, i c_a c_b.
\end{eqnarray}

Integrating out the strongly coupled modes $c_a$ and $c_b$ generates an effective coupling between the remaining modes $b_{1x}$ and $c_b$: 
\begin{equation}
H_\mathrm{eff} = -\frac{h_x J_x}{J_z} \, u_{1a} u_{ab} \, i b_1^x c_b. 
\end{equation}

After repeating the process enough times, we generate an effective coupling between the dangling Majorana modes $b_{1x}$ and $b_{2x}$:

\begin{equation}
H_\mathrm{eff} = \frac{h_x^2}{J_z}  
	\sum_\mathrm{paths} \frac{J_x^{n_x} J_y^{n_y}}{J_z^{n_x+n_y}}  
		\, i U_{12} b_1^x b_2^x.
\label{eq:b-Heff}
\end{equation}
The effective interaction depends on the fermion parity (\ref{eq:pi12}), confirming our guess that it is a physical observable.

The sum in Eq.~(\ref{eq:b-Heff}) is taken over paths $1a\ldots qr2$ with $n_\alpha$ links of type $\alpha$. Paths must alternate between weak and strong bonds and thus can propagate only upward or downward, staying within overlapping 60-degree wedges with vertices at the dislocations, Fig.~\ref{fig:wedges}. This coupling only exists for dislocations on different sublattices. For $J_x/J_z = J_y/J_z = j \ll 1$, the energy splitting induced by the potential depends on the length $L$ of a path between dislocations as $j^{(L-1)/2}$. A similar anisotropic interaction was found by \textcite{Willans} between vacancy-induced magnetic moments.

\subsection{Unpaired Majorana modes and the net fermion parity}

The presence of Majorana modes on dislocations is expected to affect the budget of fermion parity. The transmutation of the flux type upon crossing the branch cut [Fig.~\ref{fig:fig2}(b)] is analogous to the case of a torus with an odd number of rows [Fig.~\ref{eq:fig-torus}(b)]. One might therefore anticipate that the combined parity will involve, in addition to $\pi_\psi$ and $\pi_{e \times m}$, the gauge string $U_{12}$ (\ref{eq:U12}) along the cut. However, this time the gauge string has ends. To make this object gauge-invariant, we must cap its ends with Majorana operators, thereby transforming the gauge string $U_{12}$ into the parity of the Majorana modes $\pi_{12}$ (\ref{eq:pi12}). 

This turns out to be the correct guess. An evaluation of the product $W_m \pi_\psi$ in terms of spin operators in the presence of a dislocation pair [Fig.~\ref{fig:fig2}(b)] yields
\[
W_m \pi_\psi = \sigma_1^y \sigma_a^x \ldots \sigma_q^y \sigma_r^x \sigma_2^z,
\]
which agrees with the expression for $\pi_{12}$ (\ref{eq:pi12-spins}). Upon replacing $W_m$ with fermion parity $\pi_{e \times m}$ we obtain conservation of combined parity,
\begin{equation}
\pi_{\psi} \, \pi_{e \times m} \, \pi_{12} = 1. 
\label{eq:pi-psi-em-12}
\end{equation}

When a flux crosses the gauge string between the two dislocations, two quantities in Eq.~(\ref{eq:pi-psi-em-12}) switch signs: the conversion of flux type alters $\pi_{e \times m}$, while the change of sign of the gauge string $U_{12}$ alters $\pi_{12}$. The net fermion parity remains unchanged. Thus the presence of the nonlocal fermion mode is fully consistent with the constraints of the physical subspace.

\subsection{Trivial dislocations} 

A trivial dislocation has a core of the same shape as its twist counterpart. However, thanks to a different orientation of the octagon core, the missing bond at its cusp is strong, Fig.~\ref{fig:fig2}(c). The missing bond leaves a dangling $b^z$ Majorana mode at the cusp. In addition, a trivial dislocation has a second free Majorana mode of the $c$ type. If the weak bonds are completely switched off, $J_x = J_y = 0$, the additional zero mode is the $c$ fermion at the cusp. At nonzero $J_x$ and $J_y$, but still in the gapped phase ($J_x + J_y < J_z$), the zero mode is a superposition of $c$ fermions in the vicinity of the cusp, as in the case of a vacancy.\cite{Willans} The two zero modes can be combined to form a local, gauge-invariant (and thus physical) degree of freedom that acts like a free magnetic moment. Its gyromagnetic tensor $g$ has only one nonzero component $g^{zz}$. 

A trivial dislocation thus behaves very much like a vacancy. Its unpaired Majorana mode $b^z$ is susceptible to local noise due to the presence of a second unpaired Majorana mode of the $c$ type that it could couple to. The additional mode is absent in a twist dislocation, so its unpaired Majorana mode is robust.

\subsection{Synthetic dislocations}
\label{sec:synthetic}

One of the most anticipated potential applications of non-Abelian anyons lies in the field of topological quantum computing. We can manipulate the state of the Majorana fermion pair at the dislocations by winding a vortex around one of them, but braiding unpaired Majorana particles would be more computationally powerful. Additionally, it is desirable to be able to create effective lattice dislocations in unperturbed Kitaev honeycomb systems in a controlled manner. It turns out that both of these goals can be achieved with the use of a strong magnetic field applied along a line of spins, similarly to the mechanism that has been suggested for the $\mathbb{Z}_{N}$ rotor models.\cite{You2}

\begin{figure}
\includegraphics[width=0.99\columnwidth]{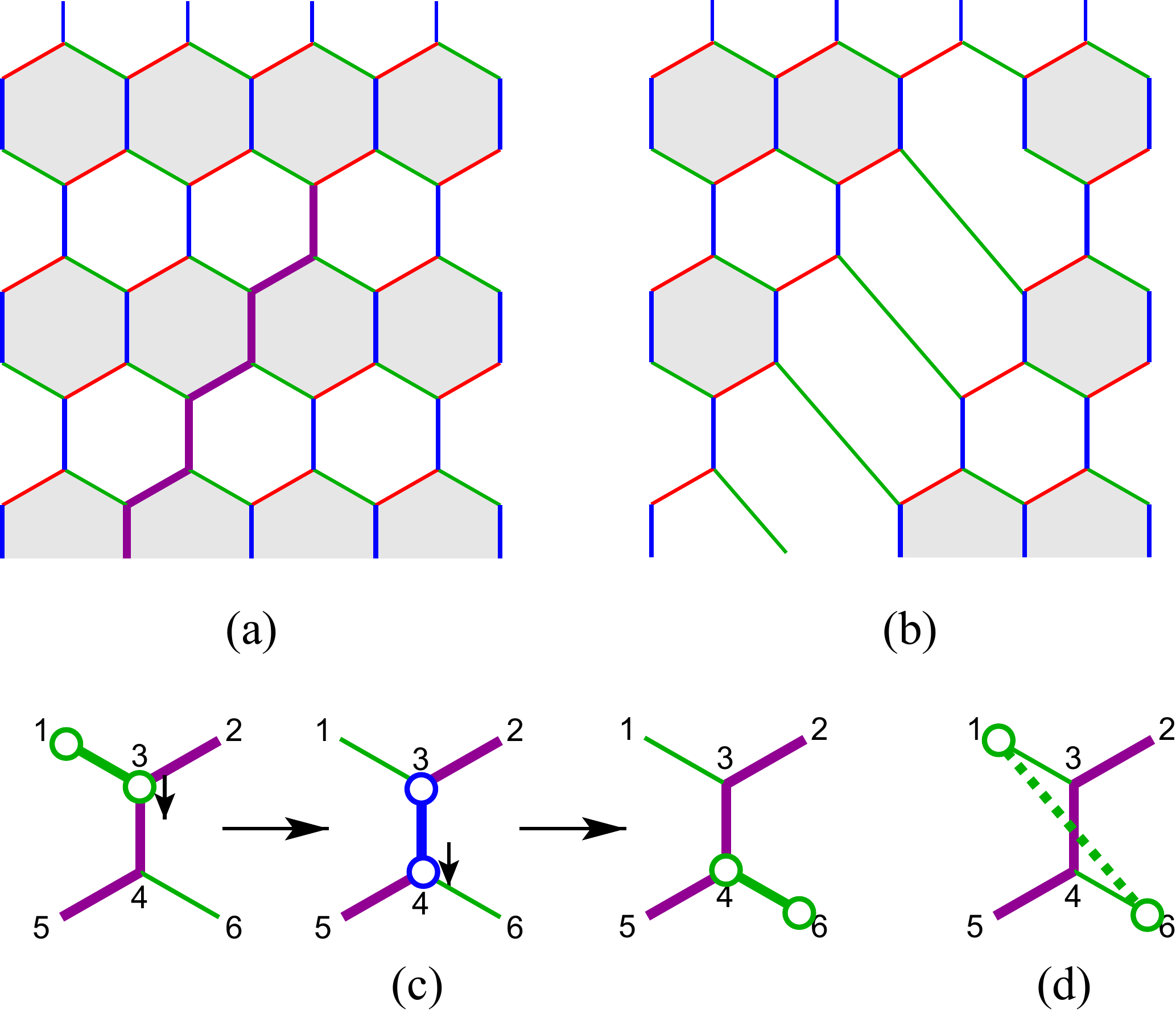}
\caption{(color) (a) Magnetic field $h_x\gg J_{\alpha}$ is applied to the spins along the purple line $C$. The original alternating $z$-$y$ flavors of the bonds are preserved. (b) The effective description of such set up results in a defect that behaves as an 8--2 dislocation. (c) In order to keep the spins aligned with the applied field, three bond terms from the Kitaev Hamiltonian must be applied together as shown. Thick green and blue bonds with open circles at the ends indicate applications of $J_{y}\sigma^{y}_{m}\sigma^{y}_{n}$ and $J_{z}\sigma^{z}_{m}\sigma^{z}_{n}$ respectively. The three operations in (c) are equivalent to connecting sites 1 and 6 directly with an effective $y$ bond (d).}
\label{fig:synthetic}
\end{figure}

Consider an unperturbed Kitaev honeycomb system in the gapped phase $J_{z} > J_{x} + J_{y}$. Let us apply a magnetic field in the $x$ direction along a zigzag line $C$ as shown in Fig. \ref{fig:synthetic}(a). In the limit where $h_x\gg J_{x},J_{y},J_{z}$ the spins located along $C$ will be aligned with $h_x$. Consider the terms in the Kitaev Hamiltonian (\ref{eq:HKitaev}) that involve those spins: all terms except for $J_{x}\sigma^{x}_{m}\sigma^{x}_{n}$ will misalign a spin with the applied field. The lowest order low energy operation then involves three bond operators forming a $y$-$z$-$y$ zigzag shown in Fig. \ref{fig:synthetic}(c). We may think of two sites at the ends of the zigzag as connected by a bond corresponding to the Hamiltonian term $\left(3J_y^2J_z/8h_x^2\right)\sigma^y_1\sigma^y_6$, and exclude sites along $C$ from the effective Hamiltonian. The $x$ bond at the cusp of the synthetic dislocation can also be omitted when $J_z\gg J_x,J_y$.

\section{5--7 dislocations}
\label{sec:57}

Fig.~\ref{fig:fig2}(d) shows a 5--7 dislocation with a Burgers vector $\mathbf B = \mathbf a_2$. Even though it has the right Burgers vector, this dislocation is not a twist. The presence of two disclinations at the core changes the orientation of $x$ and $z$ bonds along a line extending from the core. A vortex crossing the defect line in low-energy motion comes off a Burgers contour and will not change its flavor upon winding around the dislocation. Plaquette types can be globally assigned without ambiguity using low-energy vortex motion. This dislocation is trivial and is thus need not host a free Majorana mode.  

The situation changes if the strength of exchange coupling is determined by the bond's orientation, rather than its type, Fig.~\ref{fig:fig2}(e). In this case, a vortex follows a Burgers contour and changes flavor, making the dislocation a twist. One of the sites at the dislocation core has a $c$ operator weakly coupled to its neighbors. The unpaired Majorana mode is a superposition of that $c$ operator with its neighbors along two 60-degree wedges extending in both vertical directions. Its interaction with other unpaired Majorana modes is similar to that of a $b$ mode at an octagon dislocation, Eq.~(\ref{eq:b-Heff}), with two distinctions: the $c$ mode couples in both vertical directions and does not require a magnetic field for coupling. The shaded path in Fig.~\ref{fig:fig2}(d) contains a Majorana chain with regular alternation and a gapped excitation spectrum. The one in Fig.~\ref{fig:fig2}(e) has a defect---a domain wall between the two possible alternating patterns---that binds a zero mode. The fact that the  ends of the string possess free Majorana modes indicates that a lattice need not to be deformed to create twists. A topological defect such as a string  is sufficient to alter a vortex’s flavor upon braiding at its ends. This zero mode has a one-dimensional analog in Kitaev's Majorana chain with alternating weak and strong bonds \cite{Kitaev2001} and in other fermionic models. \cite{JackiwRebbi, SuSchriefferHeeger} 

\section{Flux binding by dislocations}

In addition to the transmutation of vortex flavor, lattice dislocations have interesting local properties, including binding of the flux in the ground state in the case of 8--2 dislocations.  In order to investigate this further, we first introduce a method we used to calculate the energy cost of having a flux through a dislocation, as well as the interactions between the unpaired Majorana modes when they are coupled to the system through local perturbations. 

\subsection{Diagrammatic perturbation theory}
\label{sec:diagrammatic-perturbation}

The energy of the ground state (\ref{eq:E-zero-point}) can be viewed as the energy of fermion zero-point motion. This gives a convenient starting point for developing a perturbation theory. Working with spin variables, one has to rely on the Rayleigh-Schr{\"o}dinger perturbation theory, which is rather cumbersome at high orders and in the presence of degeneracy. Switching to a fermion representation allows us to use a more economical language of Feynman diagrams. This method has the added advantage of making the $Z_2$ gauge structure of the problem manifest. 

\begin{figure}
\includegraphics[width=0.85\columnwidth]{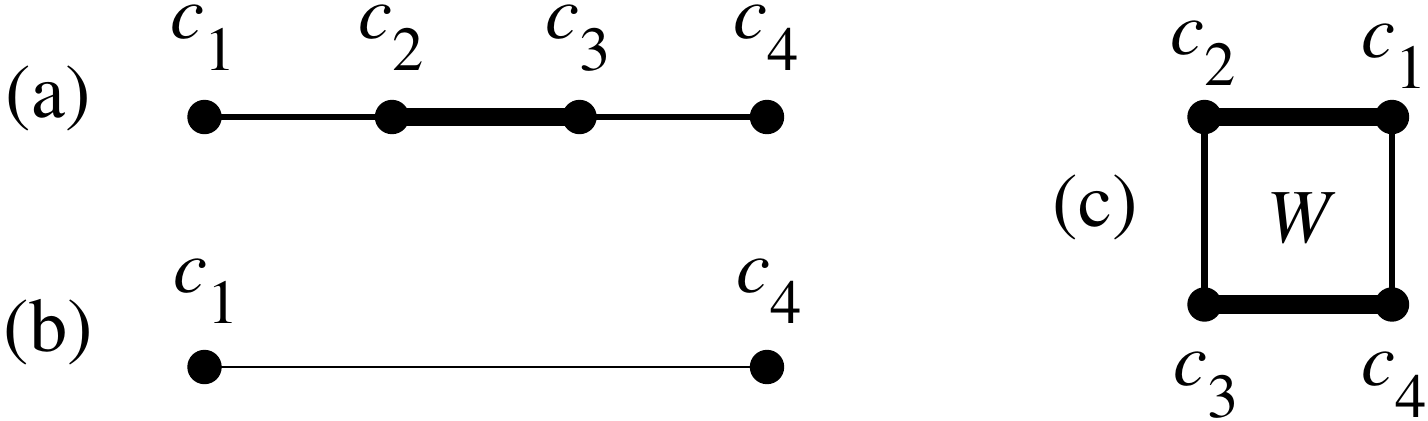}
\caption{Four coupled Majorana modes. Thick and thin lines represent strong and weak couplings. Integrating out high-energy fermions $c_2$ and $c_3$ in (a) produces an effective coupling between the remaining modes $c_1$ and $c_4$ (b). (c) Integrating out all fermion modes yields an energy correction that depends on the $Z_2$ flux $W = u_{12} u_{23} u_{34} u_{41}$.}
\label{fig:4-Majoranas}
\end{figure}

We first illustrate the idea on simple examples with four Majorana fermions $c_1$ through $c_4$ (Fig.~\ref{fig:4-Majoranas}). 

\subsubsection{Integrating out high-energy fermions.} 
\label{sec:preliminaries-perturbation-example-1}

The first case we consider is shown in Fig.~\ref{fig:4-Majoranas}(a), where modes $c_2$ and $c_3$ are strongly coupled to each other and weakly coupled to modes $c_1$ and $c_4$: 
\begin{equation}
H = i \Lambda c_2 c_3/2 + i \lambda (c_1 c_2 + c_3 c_4)/2, 
\quad \lambda \ll \Lambda.  
\end{equation}
This Hamiltonian can be easily diagonalized following the standard procedure,\cite{Kitaev2006} to obtain two (complex) fermion modes with energies $\epsilon_1 = \Lambda/2 + \sqrt{(\Lambda/2)^2 + \lambda^2} \approx \Lambda$ and $\epsilon_2 = - \Lambda/2 + \sqrt{(\Lambda/2)^2 + \lambda^2} \approx \lambda^2/\Lambda$. The high-energy mode $\epsilon_1$ is associated primarily with fermions $c_2$ and $c_3$, whereas the low-energy mode $\epsilon_2$ with $c_1$ and $c_4$. The low-energy subspace is described by an effective Hamiltonian
\begin{equation}
H_\mathrm{eff} = i \epsilon_2 c_1 c_4/2. 
\end{equation}

It is convenient to view this procedure as integrating out the high-energy fermions $c_2$ and $c_3$ and generating a new coupling $\epsilon_2 = \lambda^2/\Lambda$ between the remaining fermions $c_1$ and $c_4$, as indicated in Fig.~\ref{fig:4-Majoranas}(b).

\subsubsection{Flux dependence of zero-point energy.} 
\label{sec:preliminaries-perturbation-example-2}

This time, strong coupling exists between Majorana modes $c_1$ and $c_2$ and between $c_3$ and $c_4$ [Fig.~\ref{fig:4-Majoranas}(c)]: 
\begin{equation}
H = i \Lambda(u_{12} c_1 c_2 + u_{34} c_3 c_4)/2 + i \lambda (u_{23} c_2 c_3 + u_{41} c_4 c_1)/2. 
\end{equation}
We have added $Z_2$ gauge variables to see how the fermion zero-point energy depends on the flux $W = u_{12} u_{23} u_{34} u_{41}$. Diagonalization yields fermion energies $\epsilon_{1,2} = \Lambda \pm \lambda$ for $W = +1$ and $\sqrt{\Lambda^2 + \lambda^2}$ (doubly degenerate) for $W=-1$. The vacuum energy (\ref{eq:E-zero-point}) for the two flux values is
\begin{eqnarray}
W = +1: 
\quad 
E_0 &=& -\Lambda,
\label{eq:E0-square-plaquette}
\\
W = -1: 
\quad
E_0 &=& - \sqrt{\Lambda^2 + \lambda^2} \approx - \Lambda - \lambda^2/2\Lambda.
\nonumber
\end{eqnarray}
Adding a $\pi$ flux to a plaquette with four sites lowers the fermion zero-point energy.  

From the standpoint of perturbation theory, $-\Lambda$ is the energy of the system with the weak bonds switched off. Flux dependence of the zero-point energy comes at the second order in $\lambda$. Once again, these corrections arise from integrating out strongly paired fermions (in this case, all four). They can be computed systematically by applying the following diagrammatic rules derived in the Appendix.

\subsubsection{Diagrammatic rules}
\label{sec:preliminaries-perturbation-diagrammatic-rules}

\begin{enumerate}

\item Construct all possible directed closed paths using weak links. Treat strong links as connections that complete these paths. 

\item Compute the amplitude of a path by multiplying the following factors. Each weak link $(mn)$ contributes a factor $\lambda u_{mn}$. Each strong link $[mn]$ contributes a factor $\Lambda u_{mn}/(\omega^2 + \Lambda^2)$. Each strong link attached to the path with only site contributes a factor $\omega/(\omega^2 + \Lambda^2)$. Give an overall factor $1/2$. Integrate over the  frequency range $-\infty < \omega < +\infty$. 
 
\item Sum over all distinct closed paths. The reverse of a non-self-retracing path is a distinct path. 

\end{enumerate}

In the second example, Fig.~\ref{fig:4-Majoranas}(c), at the order $\lambda^2$ we have two self-retracing paths, (23)(32) and (14)(41). The former contributes 
\begin{equation}
\int \frac{d\omega}{2\pi} \, \frac{1}{2} \, 
	\lambda u_{23} \frac{\omega}{\omega^2 + \Lambda^2} \lambda u_{32} \frac{\omega}{\omega^2 + \Lambda^2}
	= - \frac{\lambda^2}{8\Lambda},
\end{equation}
and so does the latter. There are also two non-self-retracing paths at this order, (23)[34](41)[12] and its reverse, (14)[43](32)[21]. The former contributes 
\begin{equation}
\int \frac{d\omega}{2\pi} \, \frac{1}{2} \,
	\lambda u_{23} \frac{\Lambda u_{34}}{\omega^2 + \Lambda^2} \lambda u_{41} \frac{\Lambda u_{12}}{\omega^2 + \Lambda^2}
	= \frac{\lambda^2}{8\Lambda} W,
\end{equation}
and so does the latter. The net correction to the zero-point energy at order $\lambda^2$ is $(W-1)\lambda^2/4\Lambda$, in agreement with Eq.~(\ref{eq:E0-square-plaquette}).

Several remarks are in order. 
\begin{enumerate}
\item Flux-dependent contributions only come from non-self-retracing paths. Therefore, the lowest order at which a flux-dependent correction to the energy of a plaquette with $m$ weak links appears is $\lambda^m$. 
\item A path of length $n$ and its reverse contribute amplitudes that differ by a factor $(-1)^n$. Therefore, closed paths with an odd number of links do not contribute to energy. 
\item The amplitude from a non-self-retracing path with $n$ links is a positive number times $u_{12} u_{23} \ldots u_{n1} = (-1)^{n/2} \, W$. Therefore, a plaquette with perimeter $4n+2$, e.g., a hexagon, has $W=+1$ in the ground state, whereas a plaquette with perimeter $4n$, e.g., a square, has $W = -1$. 
\item Weak coupling constants $\lambda$ may vary from link to link and so can strong coupling constants $\Lambda$.
\end{enumerate}

\subsubsection{Energies of $Z_2$ vortices}

\begin{figure}
\includegraphics[width=0.97\columnwidth]{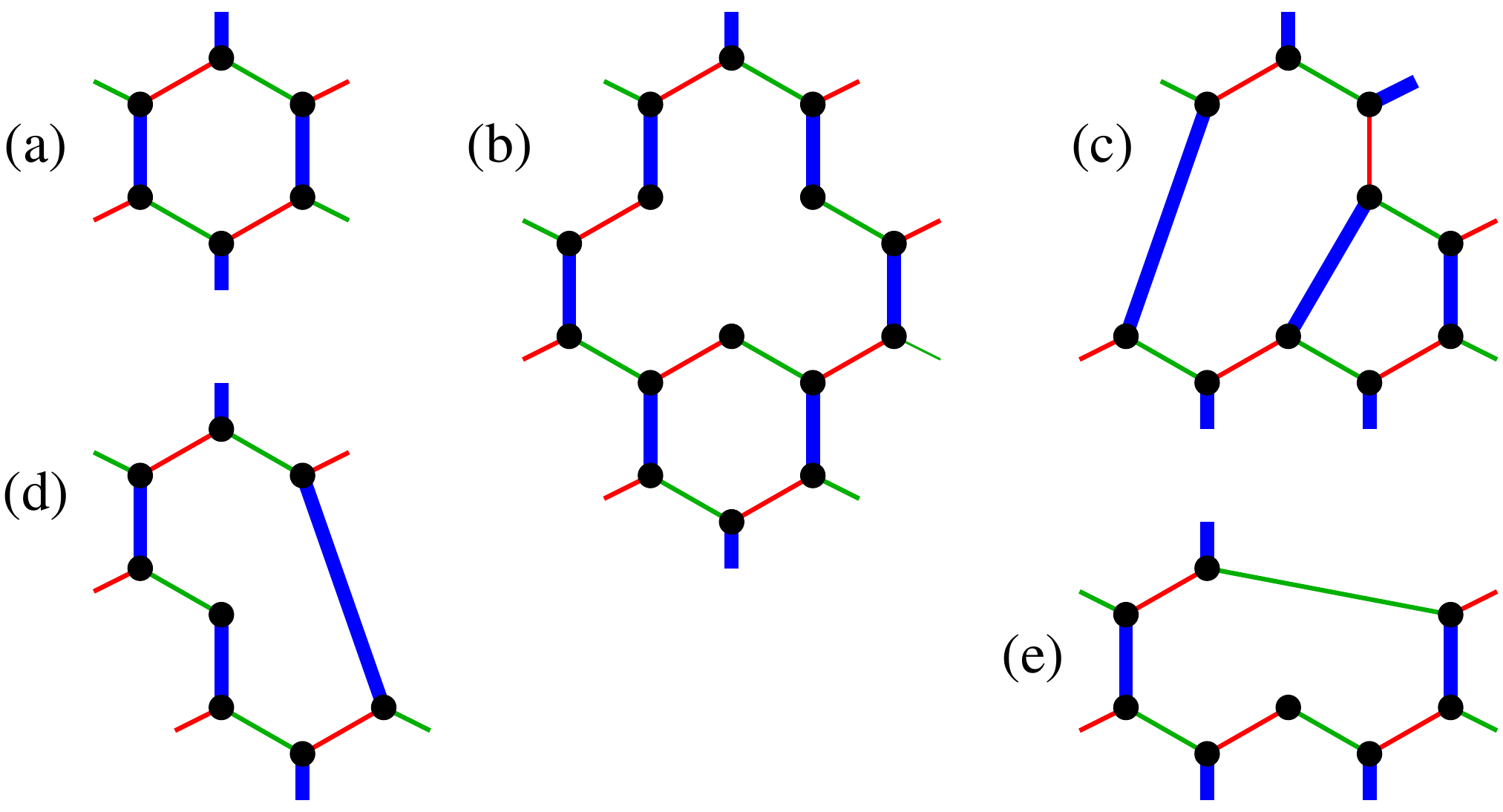}
\caption{(a) A hexagon. (b) A hexagon and a clover leaf near a vacancy. (c) A 5--7 dislocation. (d) A twist 8--2 dislocation. (e) A trivial 8--2 dislocation. }
\label{fig:flux-hexagon-cloveleaf}
\end{figure}

It is now easy to obtain the leading flux-dependent energy correction for a hexagonal plaquette in a honeycomb lattice, Fig.~\ref{fig:flux-hexagon-cloveleaf}(a). Two twin paths (the reverses of each other) contribute an energy correction at order $\lambda^4$. The paths have two weak links of strength $\lambda = 2J_x$ and two of strength $\lambda = 2 J_y$, two strong links of strength $\Lambda = 2 J_z$, and two attached strong links of strength $\Lambda = 2 J_z$. The flux-dependent energy correction at this order is  
\begin{widetext}
\begin{equation}
2 \times (-1)^3 \, W \int \frac{d\omega}{2\pi} \, \frac{1}{2} \,
	\left( 2 J_x \right)^2 
	\left( 2 J_y \right)^2 
	\left( \frac{2 J_z}{\omega^2 + (2 J_z)^2} \right)^2
	\left( \frac{\omega}{\omega^2 + (2 J_z)^2} \right)^2
	= - \frac{J_x^2 J_y^2}{16 J_z^3} W,
\end{equation}
in agreement with Kitaev.\cite{Kitaev2003}

\textcite{Willans} considered the problem of a vacancy in Kitaev's model. In effect, a vacancy removes three links. The  hexagon in Fig.~\ref{fig:flux-hexagon-cloveleaf}(b) (Fig.~2 in Ref.~\onlinecite{Willans}) is missing one of the attached strong links. Setting $\Lambda = 0$ for that link, we obtain the flux-dependent energy for such a hexagon at the fourth order,
\begin{equation}
2 \times (-1)^3 \, W \int \frac{d\omega}{2\pi} \, \frac{1}{2} \,
	\left( 2 J_x \right)^2 
	\left( 2 J_y \right)^2 
	\left( \frac{2 J_z}{\omega^2 + (2 J_z)^2} \right)^2
	\frac{\omega}{\omega^2 + (2 J_z)^2} 
	\frac{\omega}{\omega^2 + 0^2}
	= - \frac{3J_x^2 J_y^2}{8 J_z^3} W,
\end{equation}
which correctly reproduces their result. 

Finally, we evaluate the flux-dependent energy for the clover-shaped plaquette obtained by merging the three hexagons around a vacancy, Fig.~\ref{fig:flux-hexagon-cloveleaf}(b) (Fig.~2 in Ref.~\onlinecite{Willans}). The leading flux-dependent energy correction comes from two twin paths with perimeter 12 that include 4 weak links with $\lambda = 2 J_x$, four weak links with $\lambda = 2 J_y$, and four strong links with $\Lambda = 2 J_z$; 3 attached strong links have $\Lambda = 2 J_z$, and one strong link is missing, $\Lambda = 0$. They add energy 
\begin{equation}
2 \times (-1)^6 \, W \int \frac{d\omega}{2\pi} \, \frac{1}{2} \,
	\left( 2 J_x \right)^4 
	\left( 2 J_y \right)^4 
	\left( \frac{2 J_z}{\omega^2 + (2 J_z)^2} \right)^4
	\left( \frac{\omega}{\omega^2 + (2 J_z)^2} \right)^3
	\frac{\omega}{\omega^2 + 0^2}
	= \frac{21 J_x^4 J_y^4}{2^{10} J_z^7} W.
\end{equation}
Deriving this result in the language of spin variables requires a computation of $8! = 80640$ different terms in perturbation theory.\cite{Willans} The diagrammatic method requires considerably less effort. 
\end{widetext}

\subsection{8--2 dislocations}

We next show that 8--2 dislocations bind a $Z_2$ vortex. To that end, we compute the leading-order dependence to the fermion zero-point energy on the flux $W$ through the octagonal plaquette at the core of the dislocation using the diagrammatic method described in Sec.~\ref{sec:diagrammatic-perturbation}. For a twist dislocation, Fig.~\ref{fig:flux-hexagon-cloveleaf}(d), the two shortest closed paths follow the perimeter of the octagonal plaquette (clockwise and counterclockwise). They contain 3 weak links of strength $\lambda=2J_x$, 2 weak links with $\lambda=2J_y$, and 3 strong links with $\Lambda=2J_z$. 2 strong links ($\Lambda=2J_z$) are adjacent to this path. These two twin paths give the following contribution to zero-point energy:
\begin{widetext}
\begin{equation}
2\times (-1)^4 W
	\int_{-\infty}^{\infty}\frac{d\omega}{2\pi} 
		\frac{1}{2} (2J_x)^3 (2J_y)^2 
		\left(\frac{2J_z}{\omega^2+(2J_z)^2}\right)^3
		\left(\frac{\omega}{\omega^2+(2J_z)^2}\right)^2 
= \frac{5 J_x^3 J_y^2}{128 J_z^4} W.
\label{eq:W-dependence-twist-dislocation}
\end{equation}
The energy is lowered if a $Z_2$ vortex is present, $W=-1$. 

A similar calculation for a trivial dislocation, Fig.~\ref{fig:flux-hexagon-cloveleaf}(e), yields the leading-order energy correction dependent on the $Z_2$ flux 
\begin{equation}
2\times (-1)^4 W 
	\int_{-\infty}^{\infty} \frac{d\omega}{2\pi} 
		\frac{1}{2} (2J_x)^3 (2J_y)^3 
		\left(\frac{2J_z}{\omega^2+(2J_z)^2}\right)^2
		\left(\frac{\omega}{\omega^2+(2J_z)^2}\right)^3
		\frac{\omega}{\omega^2+0^2}
	= \frac{5 J_x^3J_y^3}{128 J_z^5} W.
\label{eq:W-dependence-trivial-dislocation}
\end{equation}
\end{widetext}
Again, the dislocation binds a vortex.

\subsection{5--7 dislocations}

It is easy to show that a dislocation with a 5--7 core does not bind a $Z_2$ vortex. Closed paths around the pentagon plaquette have an odd perimeter and thus do not contribute to the zero-point energy, as explained in Sec.~\ref{sec:preliminaries-perturbation-diagrammatic-rules}. The same applies to paths going around the heptagonal plaquette. The shortest loop whose zero-point energy contribution depends on the flux is the path of length 10 going around both the pentagon and the heptagon, Fig.~\ref{fig:flux-hexagon-cloveleaf}(c). This path and its reverse contribute 
\begin{widetext}
\begin{equation}
2\times (-1)^5 W 
	\int_{-\infty}^{\infty} \frac{d\omega}{2\pi}
		\frac{1}{2} (2J_x)^4 (2J_y)^4 
		\left(\frac{2J_z}{\omega^2+(2J_z)^2}\right)^2
		\left(\frac{\omega}{\omega^2+(2J_z)^2}\right)^6
	= -\frac{5}{2048}\frac{J_x^4J_y^4}{J_z^7}W
\label{eq:delta57}
\end{equation}
\end{widetext}
to the zero-point energy. The energy is minimized by setting $W=+1$ so that 5--7 dislocations do not bind vortices.

\section{Discussion}
\label{sec:discussion}

In his seminal paper on the honeycomb spin model, \textcite{Kitaev2006} posited the possibility of having unpaired Majorana fermion modes in the Abelian phase of the model. In our work we have demonstrated the existence of such modes explicitly. We showed that unpaired Majorana fermions are found in the presence of so-called twist defects\cite{Bombin} associated with the symmetry of the Abelian phase under the exchange of $e$ and $m$-type fluxes. When a flux winds around a twist defect, it changes its type. In the meantime, a non-local fermion associated with the unpaired Majorana modes at the two twists is created or annihilated. We verified that the total fermionic parity is conserved in this process and that the non-local fermion mode is physical.

The twist defects that we study in this work are realized in certain kinds of lattice dislocations. Whether a dislocation is a twist depends on its Burgers vector as well as on its internal structure. The non-local fermion, composed of the two unpaired Majorana modes localized at dislocations and a gauge string between them, is a zero mode. We have shown that separating the two dislocations in space would make the zero mode stable with respect to any local perturbations. We did so by using an applied magnetic field as a perturbation, and found that indeed the splitting of the zero mode decays exponentially with the distance between dislocations. It is interesting to note that for 8--2 dislocations, only those with the reduced coordination number sites located on different sublattices and within each other's 60 degree wedges, are able to interact. A similar result was obtained for the vacancy problem\cite{Willans}.

Our study of 5--7 dislocations suggested that it is not necessary to introduce lattice dislocations into the system in order for it to have twists. Instead, one could start with a lattice without dislocations and introduce a string defect, along which alternating weak and strong bonds are interchanged, Fig.~\ref{fig:fig2}(f). The ends of the string act as twists and possess free Majorana modes of the $c$ type. Additionally, inspired by the work done for the toric code,\cite{You2} we considered another type of twist defect that does not require altering the geometry of the lattice: synthetic dislocations created via the application of a magnetic field along a line of sites.

\section*{Acknowledgments}

We thank L. Balents, J.T. Chalker, R. Moessner, M. Oshikawa, S. Parameswaran, Y. Wan, X.-G. Wen, and H. Yao for useful discussions. We acknowledge the hospitality of the Kavli Institute for Theoretical Physics and of the Aspen Center for Physics, where part of this work was done. This work was supported in part by the US Department of Energy, Office of Basic Energy Sciences, Division of Materials Sciences and Engineering under Grant No. DE-FG02-08ER46544 (JHU), by the US National Science Foundation under Grants No. PHY-1066293 (ACP) and PHY-1125915 (KITP), by Fondecyt under Grant No. 11121397 and Conicyt under Grant No. 79112004 (AIU).  O. P. gratefully acknowledges the support of the Max Planck Society and the Alexander von Humboldt Foundation.

\appendix

\section{Derivation of the diagrammatic perturbation theory}
\label{app:diagrams}

In this section we derive the diagrammatic perturbation theory for Majorana fermions in the limit where one of the coupling constants dominates, e.g., $J_z \gg J_x, \, J_y$. Switching off the weak couplings, $J_x = J_y = 0$, leaves all spins coupled pairwise, with the ground-state energy $-J_z$ contributed by each pair of so coupled spins. The perturbation theory computes corrections to this value due to small couplings $J_x$ and $J_y$. We have found that the formalism of fermion path integrals \cite{Altland-Simons} provides a much simpler and intuitive way to evaluate higher-order corrections than the standard Rayleigh-Schr{\"o}dinger perturbation theory applied to spin variables. \cite{Kitaev2006, Willans} In the fermionic formalism, the quantity of interest is the zero-point energy of the $c$ fermion modes (\ref{eq:E-zero-point}).

\subsection{Grassmann variables}

For two Grassmann variables $a$ and $\bar{a}$ with Gaussian action $S = K \bar{a} a$,
\begin{eqnarray}
&&\int d\bar{a} \, da \, \exp{(- K \bar{a} a)} = K,
\\
&&\langle a \bar{a} \rangle 
	\equiv \frac{\int d\bar{a} \, da \, a \bar{a} \exp{(- K \bar{a} a)}}{\int d\bar{a} \, da \, \exp{(- K \bar{a} a)}}
	= 1/K.
\nonumber
\end{eqnarray}
For several pairs $\{a_m, \bar{a}_m\}$ with Gaussian action $S = \bar{a} K a \equiv \bar{a}_m K_{mn} a_n$ (summation over doubly repeated indices implied), 
\begin{eqnarray}
&&\int D\bar{a} \, Da \, \exp{(- \bar{a} K a)} = \det{K},
\\
&&\langle a_m \bar{a}_n \rangle 
	\equiv \frac{\int D\bar{a} \, Da \, a_m \bar{a}_n \exp{(- \bar{a} K a)}}
				{\int D\bar{a} \, Da \, \exp{(- \bar{a} K a)}}
	= (K^{-1})_{mn}.
\nonumber
\end{eqnarray}

\subsection{Path integrals for two Majorana modes}

Consider two coupled Majorana modes $a_1$ and $a_2$ with excitation energy $\epsilon > 0$. The quantum Hamiltonian of this system is 
\begin{equation}
H = \frac{i \epsilon}{2} u_{12} a_1 a_2 = \frac{\epsilon}{2} (\psi^\dagger \psi - \psi \psi^\dagger),
\label{eq:H-2-modes}
\end{equation}
where 
\begin{equation}
\psi = \frac{a_1 + i u_{12} a_2}{2}, 
\quad
\psi^\dagger = \frac{a_1 - i u_{12} a_2}{2}. 
\end{equation}
For future reference, we have included a $Z_2$ gauge variable $u_{12} = - u_{21} = \pm 1$. 

The classical action for these two modes is
\begin{equation}
S = \frac{i}{4} \int_{t_i}^{t_f} dt 
	\left( 
		a_m \frac{d{a}_m}{dt} - a_m A_{mn} a_n
	\right),
\end{equation}
where $a_1$ and $a_2$ are anticommuting Grassmann variables and $A_{mn}$ is an antisymmetric matrix with $A_{12} = \epsilon u_{12}$; summation over doubly repeated indices $m$ and $n = 1, 2$ is implied. Variation of the action with respect to $a_1$ and $a_2$ yields classical equations of motion,
\begin{equation}
da_1/dt = \epsilon  u_{12} a_2,
\quad 
da_2/dt = \epsilon  u_{21} a_1.
\end{equation}
``Complex'' Grassmann variables $\psi = a_1 + i u_{12} a_2$ and $\bar{\psi} = a_1 - i u_{12} a_2$ satisfy the following equations:
\begin{equation}
d\psi/dt = -i \epsilon \psi. 
\quad
d\bar{\psi}/dt = i \epsilon \bar{\psi},
\end{equation}
It follows that $\psi(t) = \psi(0) e^{-i\epsilon t}$ and $\bar{\psi}(t) = \bar{\psi}(0) e^{i\epsilon t}$, as expected for annihilation and creation operators of a fermion mode with energy $\epsilon$.

The partition function $Z$ of the quantum system (\ref{eq:H-2-modes}) at inverse temperature $\beta$ can be obtained as a path integral of $e^{iS}$, where the action $S$ is computed for an imaginary time interval from 0 to $- i\beta$: 
\begin{equation}
S = \frac{i}{4} \int_0^{- i \beta} dt 
	\left( 
		a_m \frac{d{a}_m}{dt} - a_m A_{mn} a_n
	\right).
\end{equation}
It is convenient to switch to imaginary time $\tau = it$ and Euclidean action 
\begin{equation}
S_E = - i S 
	= \frac{1}{4} \int_0^{\beta} d\tau 
	\left( 
		a_m \frac{d{a}_m}{d\tau} + i a_m A_{mn} a_n
	\right)
\end{equation}
with antiperiodic boundary conditions, $a_m(\beta) = -a_m(0)$.

The partition function $Z$ is evaluated by integrating $e^{-S_E}$ over all possible paths of the Grassmann variables $a_1(\tau)$ and $a_2(\tau)$. Switching to Fourier modes with fermionic Matsubara frequencies $\omega_\nu = 2\pi \nu/\beta$, $\nu = \pm 1/2, \pm 3/2, \ldots$, 
\begin{equation}
a_m(\tau) = \frac{1}{\sqrt{\beta}} \sum_{\nu = -\infty}^\infty a_{m\nu} e^{-i \omega_\nu \tau},
\end{equation}
yields the Euclidean action 
\begin{equation}
S_E = \frac{1}{4}\sum_{\nu = -\infty}^\infty 
	a_{m,-\nu} \left(
		- \delta_{mn} i\omega_\nu + i A_{mn}
	\right) a_{n \nu}.
\end{equation}
Terms with a given $\nu$ appear in the sum twice: once for the summation index $\nu$ and once for $-\nu$. It is convenient to gather them all by restricting the sum to $\nu > 0$: 
\begin{equation}
S_E = \frac{1}{2}\sum_{\nu = 1/2}^\infty 
	a_{m,-\nu} \left(
		- \delta_{mn} i\omega_\nu + i A_{mn}
	\right) a_{n \nu}.
\end{equation}
Lastly, we rename $a_{m,-\nu}$ into $\bar{a}_{m\nu}$: 
\begin{equation}
S_E = \frac{1}{2} \sum_{\nu = 1/2}^\infty 
	\left(
		\begin{array}{cc} 
			\bar{a}_{1\nu} & \bar{a}_{2\nu}
		\end{array} 
	\right)
	\left(
		\begin{array}{cc} 
			-i \omega_\nu & i \epsilon u_{12} \\
			i \epsilon u_{12} & -i \omega_\nu
		\end{array} 
	\right)
	\left(
		\begin{array}{c} 
			a_{1\nu} \\ a_{2\nu}
		\end{array} 
	\right).
\end{equation}
Correlations for the Fourier modes are 
\begin{equation}
- \frac{i}{2} \langle a_{m\mu} \bar{a}_{n\nu} \rangle 
	= \frac{\delta_{\mu\nu}}{\omega_\nu^2 + \epsilon^2}
		\left( 
			\begin{array}{cc}
				\omega_\nu & \epsilon u_{12} \\
				\epsilon u_{21} & \omega_\nu
			\end{array}
		\right).
\end{equation}

\subsection{Perturbation theory}

The unperturbed system has Majorana modes coupled in pairs. In that limit, Majorana fermions can only propagate within the limits of a strong bond, i.e., either staying on the same site or jumping to the site connected to it by a strong bond. 

Adding a perturbations in the form of weak bonds enables Majorana modes to move around more freely. The Euclidean action can be split into two parts, $S_E^0$ expressing the action of independent strong bonds and $S_E^1$ consisting of terms $i \lambda u_{mn} a_m a_n/4$ on weak bonds. The resulting correction to the free energy can be obtained by taking the ratio of the perturbed and unperturbed partition functions: 
\[
\Delta F = - \frac{1}{\beta} \ln{\frac{Z}{Z_0}},
\]
where 
\begin{eqnarray}
\frac{Z}{Z_0} 
	&=& \frac{\int D\bar{a} Da \, \exp{(- S_E^0 - S_E^1)}}{\int D\bar{a} Da \, \exp{(- S_E^0)}} 
	= \langle \exp{(- S_E^1)} \rangle_0
\nonumber\\
	&=& \left\langle 
		1 - S_E^1 + \frac{1}{2!}(-S_E^1)^2 - \ldots
	\right\rangle_0,
\end{eqnarray}
where the averaging is done over the unperturbed Gaussian action $S_E^0$ of decoupled strong bonds. Taking the logarithm (to obtain the free energy correction) eliminates disconnected diagrams in the expansion as usual (linked cluster expansion). 

Each weak bond ($mn$) contributes to $S_E^1$ terms 
\begin{equation}
\sum_{\nu = 1/2}^\infty \frac{i \lambda}{2} [\bar{a}_{m\nu} u_{mn} a_{n \nu} + (m \leftrightarrow n)].
\end{equation}
(no sum over doubly repeated indices $m$ and $n$). The lowest-order correction occurs at order $\lambda^2$: 
\begin{widetext}
\begin{equation}
\Delta F = - \frac{1}{\beta} 
	\left \langle \frac{1}{2!}\left(
		\sum_{\nu=1/2}^\infty \frac{-i \lambda}{2} [\bar{a}_{m\nu} u_{mn} a_{n \nu} + \bar{a}_{n\nu} u_{nm} a_{m \nu}]
	\right)^2\right\rangle_0
	= - \frac{1}{\beta} \sum_{\nu=1/2}^\infty 
		\left \langle
			\frac{-i \lambda}{2} \bar{a}_{m\nu} u_{mn} a_{n \nu} \, \frac{-i \lambda}{2} \bar{a}_{n\nu} u_{nm} a_{m \nu}
		\right\rangle_0,
\end{equation}
where the factor $1/2!$ cancels against the $2!$ ways to combine the pieces. 
\end{widetext}

We now make use of Gaussian statistics and express the quartic fermion term through quadratic ones. Modes on sites $m$ and $n$ are independent when weak bonds are switched off, hence 
\begin{equation}
\Delta F = \frac{1}{\beta} \sum_{\nu=1/2}^\infty 
	\left \langle
		\frac{-i \lambda}{2} a_{m \nu} \bar{a}_{m\nu} u_{mn}  
	\right\rangle_0
	\left \langle
		\frac{-i \lambda}{2} a_{n \nu} \bar{a}_{n\nu} u_{nm}
	\right\rangle_0.	
\end{equation}
The sign has changed because we moved $a_{m\nu}$ past an odd number of Grassmann variables. Using the expressions for the onsite propagators yields 
\begin{equation}
\Delta F = \frac{1}{\beta} \sum_{\nu=1/2}^\infty 
	\lambda u_{mn} \, \frac{\omega_\nu}{\omega_\nu^2 + \epsilon^2} \,
	\lambda u_{nm} \, \frac{\omega_\nu}{\omega_\nu^2 + \epsilon^2}.
\end{equation}
In the limit of zero temperature, $\beta^{-1} \sum_\nu \to \int d\omega/2\pi$, so
\begin{equation}
\Delta F = \int_0^\infty \frac{d\omega}{2\pi} \, 
	\lambda u_{mn} \, \frac{\omega}{\omega^2 + \epsilon^2} \,
	\lambda u_{nm} \, \frac{\omega}{\omega^2 + \epsilon^2} = - \frac{\lambda^2}{8\epsilon}.
\end{equation}
As expected, the second-order correction to the ground-state energy is negative. The minus sign comes from $u_{mn} u_{nm} = -1$. 

Higher-order diagrams are constructed in the same way.

\bibliography{dislocations}

\end{document}